\documentclass[11pt]{article}
\pdfoutput=1
\usepackage{graphics,color}
\usepackage{amsmath,amssymb,amsfonts,dsfont}
\usepackage{tensor, slashed, empheq}
\usepackage[citebordercolor={.8 .8 1},urlbordercolor ={.8 .8 1},pdfstartview=FitH]{hyperref}

\newcommand{\beq}{\begin{equation}}
\newcommand{\eeq}{\end{equation}}
\newcommand{\beqn}{\begin{eqnarray}}
\newcommand{\eeqn}{\end{eqnarray}}
\newcommand{\pa}{\partial}
\newcommand{\pasls}{\pa\kern-.4em /}
\newcommand{\pasl}{\pa\kern-.55em /}
\newcommand{\Dsls}{D\kern-.52em /}
\newcommand{\Dsl}{D\kern-.65em /}
\newcommand{\qsl}{q\kern-.5em /}
\newcommand{\ksl}{k\kern-.5em /}
\newcommand{\psl}{p\kern-.45em /}
\newcommand{\bAsl}{{\rm A}\kern-.55em /}
\newcommand{\Asl}{A\kern-.55em /}
\newcommand{\Bsl}{B\kern-.55em /}
\newcommand{\epssl}{\epsilon\kern-.45em /}
\newcommand{\Fsl}{F\kern-.65em /\kern.2em}
\newcommand{\cFsl}{{\cal F}\kern-.65em /\kern.2em}
\newcommand{\Gsl}{G\kern-.65em /\kern.2em}
\newcommand{\Jsl}{J\kern-.65em /\kern.2em}
\newcommand{\Psl}{P\kern-.65em /\kern.2em}
\newcommand{\Hsl}{H\kern-.65em /\kern.2em}
\newcommand{\omsl}{\omega\kern-.65em /\kern.2em}
\newcommand{\Omsl}{\Omega\kern-.65em /\kern.2em}
\newcommand{\cAsl}{{\cal A}\kern-.55em /\kern.2em}
\newcommand{\cDsl}{{\cal D}\kern-.55em /\kern.2em}

\newcommand{\conj}{{\bf c}}
\newcommand{\Sigsl}{\Sigma\kern-.65em /\kern.2em}

\newcommand{\sfLam}{\mathsf{\Lambda}}

\def \Sigs{{\not\!{\Sigma}}~}
\numberwithin{equation}{section}

\begin{document}

\begin{center}
\vspace{1cm} { \LARGE {\bf Fermions and $D=11$ Supergravity On Squashed Sasaki-Einstein Manifolds}}

\vspace{1.1cm}
Ibrahima Bah$^{1}$, Alberto Faraggi$^{1}$, Juan I. Jottar$^{2}$,\\  Robert G. Leigh$^{2}$ and Leopoldo A. Pando Zayas$^{1}$

\vspace{0.7cm}

{\it $^{1}$Michigan Center for Theoretical Physics, Randall Laboratory of Physics, \\
     University of Michigan, Ann Arbor, MI 48109, U.S.A.}

\vspace{0.7cm}

{\it $^{2}$Department of Physics, University of Illinois,\\
      1110 W. Green Street, Urbana, IL 61801, U.S.A. }

\vspace{0.7cm}

{\tt ibbah@umich.edu, faraggi@umich.edu, jjottar2@illinois.edu, rgleigh@illinois.edu, lpandoz@umich.edu} \\

\vspace{1.5cm}

\end{center}

\begin{abstract}
\noindent
We discuss the dimensional reduction of fermionic modes in a recently found class of consistent truncations of $D=11$ supergravity compactified on squashed seven-dimensional Sasaki-Einstein manifolds. Such reductions are of interest, for example, in that they have $(2+1)$-dimensional holographic duals, and the fermionic content and their interactions with charged scalars are an important aspect of their applications. We derive the lower-dimensional equations of motion for the fermions and exhibit their couplings to the various bosonic modes present in the truncations under consideration, which most notably include charged scalar and form fields. We demonstrate that our results are consistent with the expected supersymmetric structure of the lower dimensional theory, and apply them to a specific example which is relevant to the study of $(2+1)$-dimensional holographic superconductors. 

\end{abstract}

\pagebreak

\tableofcontents

\setcounter{page}{1}
\setcounter{equation}{0}

\section{Introduction}
Over the last decade, the gauge/gravity correspondence \cite{Maldacena:1997re, Gubser:1998bc, Witten:1998qj, Aharony:1999ti} has generated an unprecedented interest in the construction of new classes of supergravity solutions. The initial efforts were naturally directed at the construction of supergravity backgrounds dual to gauge theories displaying confinement and chiral symmetry breaking \cite{Klebanov2000,Maldacena2001}. More recently, the search for supergravity backgrounds describing systems that might be relevant for condensed matter physics has considerably expanded our knowledge of classical gravity and supergravity solutions. These include hairy black holes relevant for a holographic description of superfluidity \cite{Gubser:2008px, Hartnoll:2008vx,Hartnoll:2008kx}, and both extremal and non-extremal solutions with non-relativistic asymptotic symmetry groups (see, for example, \cite{Son:2008ye, Balasubramanian:2008dm, Maldacena:2008wh, Herzog:2008wg, Adams:2008wt}).

Since we are usually interested in lower-dimensional physics, the ability to reduce ten or eleven-dimensional supergravity solutions is central. However, only in a few cases can one explicitly construct the full non-linear  Kaluza-Klein (KK) spectrum. In the context of eleven-dimensional supergravity, one of the few such examples where the full supersymmetric spectrum of the lower-dimensional theory was worked out at the non-linear level is the reduction of $D=11$ supergravity on $S^4$ obtained in \cite{Nastase:1999cb,Nastase:1999kf}. In other cases, the best that can be done is to work with a ``consistent truncation" where only a few low-energy modes are taken into account. In this context, by a consistent truncation we mean that any solution of the lower-dimensional effective theory can be uplifted to a solution of the higher dimensional theory. Typically, the intuitive way of thinking about consistent truncations includes the assumption that there is a separation of energy scales that allows one to keep only the ``light" fields emerging from the compactification, in such a way that they do not source the tower of ``heavy" modes they have decoupled from. Often  another principle at work in consistent reductions involves the truncation to chargeless modes when such charges can be defined from the isometries of the compactification manifold; for example, this is the argument behind the consistency of compactifications on tori, where the massless fields carry no charge under the U(1)$^n$ gauge symmetry. 

The kind of solutions we are interested in in this paper have as precursors some natural generalizations of Freund-Rubin solutions \cite{Freund1980} of the form  $AdS_{4}\times SE_{7}$ in $D=11$ supergravity, where $SE_{7}$ denotes a seven-dimensional Sasaki-Einstein manifold.  In \cite{Gauntlett:2007ma}, solutions of $D=11$ supergravity of this form were shown to have a consistent reduction to minimal $N=2$ gauged supergravity in four dimensions. Furthermore, a conjecture was put forward in \cite{Gauntlett:2007ma}, asserting that for any supersymmetric solution of $D = 10$ or $D = 11$ supergravity that consists of a warped product of $AdS_{d+1}$ with a Riemannian manifold $M$, there is a consistent KK truncation on $M$ resulting in a gauged supergravity theory in $(d + 1)$-dimensions.\footnote{In the context of holography, the corresponding lower-dimensional modes are dual to the supercurrent multiplet of the $d$-dimensional dual CFT.} This is a non-trivial statement, since  consistent truncations of supergravity theories are hard to come by, even in the cases where the internal manifold is a sphere. While these consistent truncations to massless modes are difficult to construct, the reductions including a finite number of charged (massive) modes were believed to be, in most cases, necessarily inconsistent. In this light, the results of \cite{Maldacena:2008wh, Herzog:2008wg, Adams:2008wt} had a quite interesting by-product: while searching for solutions of Type IIB supergravity with non-relativistic asymptotic symmetry groups, consistent five-dimensional truncations including massive bosonic modes were constructed. In particular, massive scalars arise from the breathing and squashing modes in the internal manifold, which is then a ``deformed" Sasaki-Einstein space, generalizing the case of breathing and squashing modes on spheres that had been studied in \cite{Bremer:1998zp, Liu:2000gk} (see \cite{Buchel:2006gb}, also). The corresponding truncations including massive modes in $D=11$ supergravity on squashed $SE_{7}$ manifolds were then discussed in \cite{Gauntlett:2009zw}, and we will use them as the starting point for our work. 

While the supergravity truncations we have mentioned above are interesting in their own right, they serve the dual purpose of providing an arena for testing and exploring the ideas of gauge/gravity duality, and in particular its applications to the description of strongly-coupled condensed matter systems. In fact, even though the initial holographic models of superfluids \cite{Gubser:2008px, Hartnoll:2008vx,Hartnoll:2008kx} and non-relativistic theories \cite{Son:2008ye, Balasubramanian:2008dm} were of a phenomenological (``bottom-up") nature, it soon became apparent that it was desirable to provide a stringy (``top-down") description of these systems. Indeed, a description in terms of ten or eleven-dimensional supergravity backgrounds sheds light on the existence of a consistent UV completion of the lower-dimensional effective bulk theories, while fixing various parameters that appear to be arbitrary in the bottom-up constructions. An important step in this direction was taken in \cite{Gauntlett:2009dn,Gauntlett:2009bh}, where a $(2+1)$-dimensional holographic superconductor was embedded in M-theory, the relevant feature being the presence of a complex (charged) bulk scalar field supporting the dual field theory condensate for sufficiently low temperatures of the background black hole solution, with the conformal dimensions of the dual operator matching those of the original examples \cite{Hartnoll:2008vx,Hartnoll:2008kx}. At the same time, a model for a $(3+1)$-dimensional holographic superconductor embedded in Type IIB string theory was constructed in \cite{Gubser:2009qm}.

 Some of the Type IIB truncations have been recently brought into the limelight again, and  a more complete and formal treatment of the reduction has been reported. In particular, consistent $N = 4$ truncations of Type IIB supergravity on squashed Sasaki-Einstein manifolds including massive modes have been studied in \cite{Cassani:2010uw} and \cite{Gauntlett:2010vu}, while \cite{Liu:2010sa} also extended previous truncations to gauged $N =2 $ five-dimensional supergravity to include the full bosonic sector coupled to massive modes up to the second KK level. Similarly, \cite{Skenderis:2010vz} studied holographic aspects of such reductions as well as the properties of solutions of the type $AdS_4\times \mathds{R}\times SE_5$. Issues of stability of vacua have been considered in Ref. 
\cite{Bobev:2010ib}.

It is important to realize that, with the exception of \cite{Nastase:1999cb,Nastase:1999kf}, all of the work on consistent truncations that we have mentioned so far discussed the reduction of the bosonic modes only,\footnote{In some cases (see \cite{Gauntlett:2007ma,Buchel:2006gb}, for example), fermions were considered to the extent that the lower-dimensional solutions preserving supersymmetry were shown to uplift to higher-dimensional solutions which also preserve supersymmetry.} in the hope that the consistency of the truncation of the fermionic sector is ensured by the supersymmetry of the higher-dimensional theory. In fact, this has been rigorously proven to hold in certain simple cases involving compactifications on a sphere \cite{Pope:1987ad, Cvetic:2000dm}. However, from the point of view of applications to gauge/gravity duality, it is important to know the precise form of the couplings between the various bosonic fields and their fermionic partners, inasmuch as this knowledge would allow one to address relevant questions such as the nature of fermionic correlators in the presence of superconducting condensates, that rely on how the fermionic operators of the dual theory couple to scalars. A related problem involving a superfluid $p$-wave transition was studied in \cite{Ammon:2010pg}, in the context of (3+1)-dimensional supersymmetric field theories dual to probe $D5$-branes in $AdS^{5}\times S^5$. In the case of the $(2+1)$-dimensional field theories which concern us here, some of these issues have been discussed in a bottom-up framework in \cite{Faulkner:2009am,Gubser:2009dt}. We note in particular though that in the presence of scalar excitations, the $d=4$ gravitino will mix with any other fermions (beyond the linearized approximation). The goal of the present paper is to set the stage for addressing these questions in a more systematic top-down fashion, by explicitly reducing the fermionic sector of the truncations of $D=11$ supergravity constructed in \cite{Gauntlett:2009zw,Gauntlett:2009dn,Gauntlett:2009bh}.

This paper is organized as follows. In section \ref{section:Ansatz}  we briefly review some aspects of the truncations of $D=11$ supergravity constructed in \cite{Gauntlett:2009zw,Gauntlett:2009dn,Gauntlett:2009bh} and the extension of the bosonic ansatz to include the gravitino. In section \ref{section:4d eqs} we present our main result: the four-dimensional equations of motion for the fermion modes, and the corresponding  effective four-dimensional action functional in terms of diagonal fields. In section \ref{section:N=2} we reduce the supersymmetry variation of the gravitino, and elucidate the supersymmetric structure of the four-dimensional theory by considering how the fermions fit into the supermultiplets of gauged $N=2$ supergravity in four dimensions. Thus, we explain how the reduction is embedded in the general scheme of Ref. \cite{Andrianopoli:1996cm}. In section \ref{section:Examples} we apply our results to two further truncations of interest: the minimal gauged supergravity theory in four dimensions, and the dual \cite{Gauntlett:2009dn,Gauntlett:2009bh} of the $(2+1)$-dimensional holographic superconductor. In particular, we briefly discuss the possibility of further truncating the fermionic sector which would be necessary to obtain a simpler theory of fermionic operators coupled to superconducting condensates. We conclude in section \ref{section:Conclusions}. Various conventions and useful expressions have been collected in the appendices. 

\section{$D=11$ supergravity on squashed Sasaki-Einstein manifolds}\label{section:Ansatz}
In this section we briefly review the ansatz for the bosonic fields in the consistent truncations of \cite{Gauntlett:2009zw,Gauntlett:2009dn,Gauntlett:2009bh}, and discuss the extension of this ansatz to include the gravitino. 

\subsection{The bosonic ansatz}
The Kaluza-Klein metric ansatz in the truncations of interest is given by \cite{Gauntlett:2009zw}
\begin{equation}\label{metric ansatz}
ds^2_{11}=e^{-6U(x) - V(x)}ds_{E}^2(M)+e^{2U(x)}ds^2(Y)+e^{2V(x)}\bigl(\eta +A(x)\bigr)^2\, ,
\end{equation}

\noindent where $M$ is an arbitrary ``external" four-dimensional manifold, with coordinates denoted generically by $x$ and four-dimensional Einstein-frame metric $ds_{E}^{2}(M)$, and $Y$ is an ``internal" six-dimensional K\"ahler-Einstein manifold (henceforth referred to as ``KE base") coordinatized by $y$ and possessing K\"ahler form $J$. The one-form $A$ is defined in $T^*M$ and $\eta \equiv d\chi + {\cal A}(y)$, where ${\cal A}$ is an element of $T^*Y$ satisfying $d{\cal A} \equiv {\cal F} = 2J$. For a fixed point in the external manifold, the compact coordinate $\chi$ parameterizes the fiber of a $U(1)$ bundle over $Y$, and the seven-dimensional internal manifold spanned by $(y,\chi)$ is then a squashed Sasaki-Einstein manifold, with the breathing and squashing modes parameterized by the scalars $U(x)$ and $V(x)$.\footnote{In particular, $U-V$ is the squashing mode, describing the squashing of the $U(1)$ fiber with respect to the KE base, while the breathing mode $6U + V$ modifies the overall volume of the internal manifold. When $U=V=0$, the internal manifold becomes a seven-dimensional Sasaki-Einstein manifold $SE_{7}$.} In addition to the metric, the bosonic content of $D=11$ supergravity includes a 4-form flux $\hat{F}_{4}$; the rationale behind the corresponding ansatz is the idea that the consistency of the dimensional reduction is a result of truncating the KK tower to include fields that transform as singlets only under the structure group of the KE base, which in this case corresponds to $SU(3)$. As we will discuss below, this prescription allows for an interesting spectrum in the lower dimensional theory, inasmuch as the $SU(3)$ singlets include fields that are charged under the $U(1)$ isometry generated by $\partial_{\chi}$. The globally defined K\"ahler 2-form $J = d{\cal A}/2$ and the holomorphic $(3,0)$-form $\Sigma$ that define the K\"ahler and complex structures, respectively, on the KE base $Y$ are $SU(3)$-invariant and can be used in the reduction of $\hat F_4$ to four dimensions. The $U(1)$-bundle over $Y$ is such that they satisfy\footnote{Our conventions for the various form fields are discussed in Appendix \ref{Appendix:Conventions}.} 
\begin{equation}
\Sigma \wedge \Sigma^{*} = -\frac{4i}{3}J^{3}\, ,\qquad \mbox{and}\qquad d\Sigma = 4i {\cal A}\wedge \Sigma\, .
\end{equation}
More precisely, as will be clear from the discussion to follow below, the relevant charged form $\Omega$ on the total space of the bundle that should enter the ansatz for $\hat{F}_{4}$ is given by
\begin{equation}
\Omega \equiv e^{4i\chi}\Sigma\, ,
\end{equation}
and satisfies
\begin{equation}
d\Omega = 4i\eta \wedge\Omega \, .
\end{equation}

\noindent The ansatz for $\hat{F}_{4}$ is then \cite{Gauntlett:2009zw}
\begin{align}\label{F4 ansatz}
\hat F_4={}&f \, \mbox{vol}_4+H_3\wedge (\eta + A)+H_2\wedge J+dh\wedge J\wedge (\eta + A)+2h J^2\nonumber\\
&+\left[X(\eta + A)\wedge\Omega-\frac{i}{4}\left(dX-4iAX\right)\wedge\Omega + \mbox{c.c.}\right],
\end{align}

\noindent where, as follows from the equations of motion, $f = 6e^{6W}(\epsilon+h^2+\frac{1}{3}|X|^2)$, with $\epsilon = \pm 1$ and $W(x) \equiv  -3U(x) - V(x)/2$, a notation we will use often.\footnote{The normalization of the charged scalar $X$ is related to the one in \cite{Gauntlett:2009zw} by $X=\sqrt{3}\chi$. Here, we reserve the notation $\chi$ for the fiber coordinate.} All the fields other than $(\eta, J, \Omega)$ are defined on $\sfLam^* T^{*}M$. The matter fields $X$ and $h$ are scalars, while $H_{2}$ and $H_{3}$ are 2-form and 3-form field strengths, respectively. In terms of a 1-form potential $B_{1}$ and a 2-form potential $B_{2}$, the field strengths can be written $H_3=dB_2$ and $H_2=dB_1+2B_2+hF$, and it is then easy to verify that the Bianchi identity $d\hat{F}_{4} = 0$ is satisfied. As pointed out in \cite{Gauntlett:2009zw,Gauntlett:2009dn,Gauntlett:2009bh}, when $\epsilon = +1$ the dimensionally reduced theory admits a vacuum solution with vanishing matter fields, which uplifts to an $AdS_{4}\times SE_{7}$ eleven-dimensional solution. On the other hand, by reversing the orientation in the compact manifold (i.e. $\epsilon = -1$) the corresponding vacuum is a ``skew-whiffed" $AdS_{4}\times SE_{7}$ solution, which generically does not preserve any supersymmetries, but is nevertheless perturbatively stable \cite{Duff:1984sv}.

\subsection{The gravitino ansatz}\label{subsection:gravitino ansatz}
Quite generally, we would like to decompose the gravitino using a separation of variables ansatz of the form
\begin{align}
\psi_a(x,y,\chi) = {}&\sum_I\psi_a^I(x)\otimes \eta^I(y,\chi)\\
\psi_\alpha(x,y,\chi)={}&\sum_I\lambda^I(x)\otimes \eta_\alpha^I(y,\chi)\\
\psi_f(x,y,\chi)={}&\sum_I\varphi^I(x)\otimes \eta_f^I(y,\chi)\, .
\end{align}
The relevant point to understand is how precisely to project to $SU(3)$ singlets, appropriate to the consistent truncation. The first step is to understand how $SU(3)$ acts on the spinors, which is explored fully in Appendix \ref{Appendix:singlets}.

As we have discussed, the seven-dimensional internal space is the total space of a $U(1)$ bundle over a KE base $Y$. In general, the base is not spin, and therefore spinors do not necessarily exist globally on the base. However, it is always possible to define a $Spin^{c}$ bundle globally on $Y$ (see  \cite{Martelli:2006yb}, for example), and our ``spinors" will then be sections of this bundle. The corresponding $U(1)$ generator is proportional to $\pa_\chi$, and hence $\nabla_\alpha- {\cal A}_\alpha\pa_\chi$ is the gauge connection on the $Spin^c$ bundle, where $\nabla_{\alpha}$ is the covariant derivative on $Y$. Of central importance to us in the reduction to $SU(3)$ invariants are the gauge-covariantly-constant spinors, which can be defined on any K\"ahler manifold \cite{Hitchin19741} and thus satisfy in the present context
\beq\label{gauge cov const spinor eq}
(\nabla_\alpha- {\cal A}_\alpha\pa_\chi)\varepsilon(y,\chi)=0\, ,
\eeq
where
\begin{equation}\label{gauge cov const spinor}
\varepsilon(y,\chi)=\varepsilon(y) e^{ie\chi}
\end{equation}
for fixed ``charge" $e$. Their existence is independent of the metric on the total space of the bundle. Thus, in our discussion, solutions to \eqref{gauge cov const spinor eq} are supposed to exist, and indeed as we will see shortly they must exist in numbers sufficient to give $N=2$ supersymmetric structure in $d=4$.

Our next task is to determine the values of the charge $e$ occurring in \eqref{gauge cov const spinor}. We will do so for a general KE manifold $Y$ of real dimension $d_{b}$. Following \cite{Pope:1984jj,Gibbons:2002th}, we start by examining the integrability condition\footnote{Our Clifford algebra conventions are detailed in Appendix \ref{Appendix:Conventions}.}
\begin{align}
[\nabla_\beta,\nabla_\alpha]\varepsilon = \frac14 (R_{\delta\gamma})_{\beta\alpha}\Gamma^{\delta\gamma}\varepsilon\, .
\end{align}
The key feature is that internal gauge curvature is equal to the K\"ahler form, ${\cal F}=2J$. Given the assumption (\ref{gauge cov const spinor}) that $\nabla_\alpha\varepsilon=ie{\cal A}_\alpha\varepsilon$, we find
\begin{align}
[\nabla_\beta,\nabla_\alpha]\varepsilon &=-ie{\cal F}_{\alpha\beta}\varepsilon =-2ieJ_{\alpha\beta}\, ,
\end{align}

\noindent and hence
\beq
\frac14 (R_{\delta\gamma})_{\beta\alpha}J^{\beta\alpha}\Gamma^{\delta\gamma}\varepsilon=-2ie J_{\alpha\beta}J^{\beta\alpha}\varepsilon = 2ied_{b}\, \varepsilon\, .
\eeq

\noindent Since Y is an Einstein manifold, the Ricci form satisfies
\beq
Ric = \frac14 (R_{\delta\gamma})_{\beta\alpha}J^{\beta\alpha}e^\delta\wedge e^\gamma = (d_{b} + 2)J\, ,
\eeq

\noindent and we then conclude
\beq
Q\varepsilon \equiv -iJ_{\alpha\beta}\Gamma^{\alpha\beta}\varepsilon = \frac{4ed_{b}}{d_{b}+2} \varepsilon\, .
\eeq

\noindent In other words, the matrix $Q=-iJ_{\alpha\beta}\Gamma^{\alpha\beta}$ on the left is (up to normalization) the U(1) charge operator.\footnote{This is explored further in Appendix \ref{Appendix:singlets}, in terms of the gravitino states.} It has maximum eigenvalues $\pm d_{b}$, and the corresponding spinors have charge 
\beq
e=\pm \frac{d_{b} + 2}{4}\, .
\eeq

\noindent These two spinors are charge conjugates of one another, and we will henceforth denote them by $\varepsilon_{\pm}$. By definition, they satisfy $\cFsl\varepsilon_{\pm} = iQ\varepsilon_{\pm} = \pm id_{b}\,\varepsilon_{\pm}$, where $\cFsl \equiv (1/2){\cal F}_{\alpha\beta}\Gamma^{\alpha\beta}$. As discussed in Appendix \ref{Appendix:singlets}, the spinors with maximal $Q$-charge are in fact the singlets under the structure group, and we will use them to build the reduction ansatz for the gravitino. In the case at hand $d_{b}=6$, the structure group is $SU(3)$, and $\varepsilon_{+}$ and $\varepsilon_{-}$ transform in the ${\bf 4}$ and ${\bf \bar 4}$ of $\mbox{Spin}(6) \simeq SU(4)$, respectively, so they have opposite six-dimensional chirality:
\beq
\gamma_7\varepsilon_\pm = \pm\varepsilon_\pm\, .
\eeq

Incidentally, we can now understand why it is that $\Omega = e^{4i\chi}\Sigma$ enters the 4-form flux ansatz: defining $\Sigs=\frac{1}{3!}\Sigma_{\alpha\beta\gamma}\Gamma^{\alpha\beta\gamma}$, we can compute $[Q,\Sigs]=12\Sigs$. This means that $\Sigma$ carries charge $e_{\Sigma}=4$. Since the $Q$ charge is realized in the spinors through their $\chi$-dependence, for the holomorphic form we are lead to define $\Omega=e^{4i\chi}\Sigma$, with $\Sigma$ given by \eqref{defSigma}.

We are now in position to write the reduction ansatz for the gravitino. Taking into account the eleven-dimensional Majorana condition on the gravitino, and dropping all the $SU(3)$ representations other than the singlets, we take
\begin{align}\label{gravitino ansatz 1}
\Psi_\alpha(x,y,\chi) &= \lambda(x)\otimes \gamma_\alpha\,\varepsilon_+(y) e^{2i\chi}\\
\Psi_{\bar\alpha}(x,y,\chi) &=- \lambda^\conj(x)\otimes \gamma_{\bar\alpha}\,\varepsilon_-(y) e^{-2i\chi}\\
\Psi_f(x,y,\chi) &= \varphi(x)\otimes\varepsilon_+(y) e^{2i\chi}+\varphi^\conj(x)\otimes\varepsilon_-(y) e^{-2i\chi}\\
\Psi_a(x,y,\chi) &= \psi_a(x)\otimes\varepsilon_+(y) e^{2i\chi}+ \psi^\conj_a(x)\otimes\varepsilon_-(y) e^{-2i\chi}\, ,\label{gravitino ansatz 4}
\end{align}

\noindent where $\varphi, \lambda$ and $\psi_a$ are four-dimensional Dirac spinors on $M$, the superscript $\conj$ denotes charge conjugation,\footnote{Our charge conjugation conventions are summarized in section \ref{Appendix:ChargeConjugation}.} and we have used the complex basis introduced in \ref{Appendix:Conventions-Fluxes} for the KE base directions ($\alpha, \bar{\alpha} = 1,2,3$). Notice that all of these modes are annihilated by the gauge-covariant derivative on $Y$. Equations \eqref{gravitino ansatz 1}-\eqref{gravitino ansatz 4} provide the starting point for the dimensional reduction of the $D=11$ supergravity equations of motion down to $d=4$. 

\section{Four-dimensional equations of motion and effective action}\label{section:4d eqs}
The $D=11$ equation of motion for the gravitino is 
\begin{equation}\label{eleven-d gravitino eq}
\Gamma^{ABC}\hat{D}_B\hat\Psi_C + \frac14\frac{1}{4!}\left[\Gamma^{ADEFGC}F_{DEFG}+12\Gamma^{DE}{F^{AC}}_{DE}\right]\hat\Psi_C = 0\, .
\end{equation}
In this paper, we will consider only effects linear in the fermion fields in the equations of motion. Consequently, we will not derive the four-fermion (current-current) couplings that are certainly present in the 4-$d$ Lagrangian. These can be obtained using the same methods that we will develop here, and it would be interesting to do so, as they might be relevant for holographic applications. In Section \ref{section:N=2}, we will show that all of our results fit into the expected $d=4$ $N=2$ gauged supergravity, and so the four fermion terms could also be derived by evaluating the known expressions.

\noindent The spin connection and our conventions for the Clifford algebra and the various form fields can be found in Appendix \ref{Appendix:Conventions}. Below, we write down the effective four-dimensional equations of motion for the fermion modes $\lambda, \varphi, \psi_{a}$ on $M$ (and their charge conjugates). We then perform a field redefinition in order to write the kinetic terms in diagonal form, and present our main result: the effective four-dimensional action functional for the diagonal fermion fields. The equations of motion that follow from this action have been written explicitly in appendix \ref{Appendix:eom}.

\subsection{Reduction of covariant derivatives}
\newcommand{\de}{d_{e}}
\newcommand{\db}{d_{b}}

We make use of the gravitino ansatz discussed in section \ref{subsection:gravitino ansatz} to reduce the eleven-dimensional covariant derivatives. In what follows, we will project the various expressions to the terms proportional to the positive chirality spinor $\varepsilon_{+}$, and drop the overall factor $e^{2i\chi}$. The $\varepsilon_-e^{-2i\chi}$ contributions are the charge conjugates of the expressions that we will write and thus can be easily resurrected.

Reducing the component in the direction of the fiber, $\Gamma^{fAB}\hat D_A\hat \Psi_B$, and denoting the resulting expression by ${\cal L}_f$, we get 
\begin{align}
e^{W}{\cal L}_f={}&\left[\gamma^{ab}D_{a}+\frac12(\pa^b W)+\gamma^{b}\pasl (V+3U)+3 ie^{W+V-2U}\gamma_5\gamma^b 
-\frac14e^{V-W}F_{da} \gamma^{ab}\gamma^d\gamma_5 \right]\psi_b\nonumber\\
&+6\left[\Dsl+\frac12\pasl (W+U-V) +\frac12e^{V-W}\Fsl\gamma_5 +\frac{3i}{2}e^{W+V-2U}\gamma_{5}\right]\gamma_5\lambda\nonumber\\
&+\left(e^{V-W}\Fsl +6ie^{W+V-2U} \right)\varphi\, , \label{LHSf}
\end{align}

\noindent where we have defined the four-dimensional gauge-covariant derivative $D_a=\nabla_a-2iA_a$. Similarly, for the piece coming from the $a$-component $\Gamma^{aAB}\hat D_A\hat \Psi_B$, which we denote by ${\cal L}_{gr}^{a}\,$, after projecting we obtain
\begin{align}
e^{W}{\cal L}_{gr}^{a}={}&
\biggl[\gamma_5\gamma^{abc}D_{b}
+\frac12(\pa_b W)\gamma_5\gamma^{abc}-i\left(2e^{W-V}+\frac32e^{W+V-2U}\right) \gamma^{ac}-\frac18e^{V-W}F_{bd}\gamma^{b}\gamma^{ac}\gamma^d\biggr] \psi_{c}
\nonumber\\
&+\left[\gamma^{ab}D_{b}+\frac12(\pa_bV)\gamma^{ab}+\frac12\pa^a(W-V)+3ie^{W+V-2U}\gamma_5 \gamma^a
+\frac14e^{V-W}\gamma_5F_{bc}\gamma^c\gamma^{ab}\right] \varphi
\nonumber\\
&
+6\biggl[\gamma^{ab}D_{b}+\frac12(\pa_b U)\gamma^{ab}+\frac12\pa^a(W-U)+i(2e^{W-V}+e^{W+V-2U})\gamma_5\gamma^a 
\nonumber\\
&\hphantom{+6\biggl[}\, -\frac18\gamma_5e^{V-W}F_{bc}\gamma^b\gamma^a\gamma^c\biggr]\lambda\, .
\end{align}

\noindent Finally, for the components in the direction of the KE base, the $SU(3)$-invariants can be extracted by contracting $\Gamma^{\alpha AB}\hat D_A\hat \Psi_B$ with $\Gamma_\alpha$. After projecting, we find
\begin{align}
e^{W}{\cal L}_{b} ={}&6\gamma_5\biggl[\gamma^{ab}D_a
+\frac12(\pa^b W)
-\frac12\gamma^b\pasl(2 W-U)+i
\left(e^{W+V-2U}+2e^{W-V}\right)\gamma_{5}\gamma^{b}\nonumber\\
&\hphantom{6\gamma_{5}\biggl[}\,+\frac{1}{8}e^{V-W}F_{da}\gamma_{5}\gamma^a\gamma^b\gamma^d\biggr]\psi_b
\nonumber\\
&
+6\left[-5\Dsl-\frac52(\pasl W)+10ie^{W-V}\gamma_5+\frac72ie^{W+V-2U}\gamma_5+\frac{5}{4}e^{V-W}\gamma_5\Fsl\right]\lambda
\nonumber\\
&
+3\left[-2\Dsl-\pasl(W+V-U)
+3ie^{W+V-2U}\gamma_5+e^{V-W}\gamma_5\Fsl \right]\varphi\, .
\end{align}

\subsection{Reduction of fluxes}
Having reduced the kinetic terms for the fermion modes, we now turn to the problem of reducing their couplings to the background $4$-form flux. More explicitly, we would like to reduce
\beq\label{right-hand side}
\frac{1}{4!}\left[\Gamma^{ADEFGC}\hat{F}_{DEFG}+12\Gamma^{DE}{\hat{F}^{AC}}_{\phantom{AC}DE}\right]\hat\Psi_C
\eeq

\noindent by using the ansatz \eqref{gravitino ansatz 1}-\eqref{gravitino ansatz 4}. As we did for the kinetic terms, here we display the expressions obtained by projecting to the terms proportional to the positive chirality spinor $\varepsilon_{+}$, and drop the overall factor $e^{2i\chi}$. 

Evaluating the component of \eqref{right-hand side} in the direction of the fiber, and denoting the corresponding expression after the projection by ${\cal R}_{f}$, we get
\begin{align}
e^{W}{\cal R}_{f} ={}&
3\left[\frac12ie^{-W-2U}H_{2\, ab}\gamma^{abc}
-\frac{1}{6}e^{-2W-V}{H_{3}^{abc}}\gamma_{ab}\gamma_5
-ie^{-2U-V}(\pa^ch)\gamma_5
-4h e^{W-4U}\gamma^{c}\right]\psi_c
\nonumber\\
&
+6\Bigl[-ife^{-3W}
+2ie^{-W-2U}\gamma_5\Hsl_{2}
-4h e^{W-4U}\gamma_5
+ie^{-2U-V}(\pasl h)\Bigr]\lambda
\nonumber\\
&
+2ie^{-3U}\gamma_5\gamma^{ab}(D_a X)\psi^\conj_b +6e^{-3U}\bigl[i(\Dsl X) -4 e^{W-V}X\gamma_5\bigr]\lambda^\conj\, .
\end{align}
We note that the terms proportional to charge conjugate spinors come about, as explained in the Appendix, because $\Omsl\epsilon_-\sim\epsilon_+$, that is $\Omsl$ is proportional to a ``total raising operator" in the Fock basis for gravitino states. We also note that the gauge-covariant derivative $D$ acts on the complex scalar $X$ as $DX = dX - 4iAX$.

\noindent Similarly, for the components in the direction of the external manifold, denoted here by ${\cal R}_{gr}^{a}$, we find
\begin{align}
e^{W}{\cal R}^{a}_{gr} ={}&
\biggl[3i(\pa_bh)e^{-2U-V}\gamma^{abc}
-\frac32e^{-W-2U}H_{2\, bd}\epsilon^{abdc}
-12he^{W-4U}\gamma_5\gamma^{ac}
\nonumber\\
&
\hphantom{\biggl[}\,+ife^{-3W}\gamma^{ac}
-e^{-2W-V}{H_3}^{acb}\gamma_b
+3ie^{-W-2U}{H_2}^{ac}\gamma_5\biggr]\psi_c
\nonumber\\
&
+3\biggl[4he^{W-4U}\gamma^a
-\frac12ie^{-W-2U}H_{2\, bc}\gamma^{abc}
+\frac16e^{-2W-V}{H_{3}}^{abc}\gamma_5\gamma_{bc}
+i(\pa^ah)e^{-2U-V}\gamma_5\biggr]\varphi
\nonumber\\
&
+6\biggl[2i(\pa_bh)e^{-2U-V}\gamma^{ab}\gamma_5
+\frac{i}{6}\epsilon^{abcd}H_{3\, bcd}e^{-2W-V}
+4he^{W-4U}\gamma^a
\nonumber\\
&\hphantom{+6\biggl[}\,
-ie^{-W-2U}H_{2\, bc}\gamma^{abc}
-ie^{-W-2U}{H_2}^{ac}\gamma_c
+i(\pa^ah)e^{-2U-V}\gamma_5\biggr]\lambda
\nonumber\\
&
+2e^{-3U}\left[-i(D_bX)\gamma^{abc}+4Xe^{W-V}\gamma_5\gamma^{ac}\right]\psi^\conj_c +2ie^{-3U}(D_bX)\gamma_5\gamma^{ab}\varphi^\conj
\nonumber\\
&
+6e^{-3U}\biggl[i\gamma_5\gamma^a(\Dsl X)
+4Xe^{W-V}\gamma^{a}\biggr]\lambda^\conj\, .
\end{align}

\noindent Next, let ${\cal R}_{b}$ denote the expression obtained by contracting the components of \eqref{right-hand side} in the KE base directions with $\Gamma_{\alpha}$ and projecting to the $\varepsilon_{+}$ sector. We then find
\begin{align}
e^{W}{\cal R}_{b}={}&
\biggl[ie^{-2W-V}H_{3;bcd}\epsilon^{abcd}\gamma_5
+6ie^{-W-2U}H_{2;bc}\gamma_5
\gamma^c\gamma^{ab}
-24he^{W-4U}\gamma_5\gamma^{a}
\nonumber\\
&
\hphantom{\biggl[}\,+6ie^{-2U-V}(\pa_bh)\left(2\gamma^{ab}-\eta^{ab}\right)\biggr]\psi_a
\nonumber\\
&
+\biggl[-6ife^{-3W}\gamma_5
+12ie^{-W-2U}\Hsl_{2}
-24he^{W-4U}
+6ie^{-2U-V}\gamma_5(\pasl h)\biggr]\varphi
\nonumber\\
&
+6\biggl[-5i fe^{-3W}\gamma_5
+5e^{-2W-V}\gamma_5\Hsl_3
+7ie^{-W-2U}\Hsl_{2} -28he^{W-4U}
\nonumber\\
&\hphantom{+6\biggl[}\,+7ie^{-2U-V}\gamma_5(\pasl h)\biggr]\lambda
+6e^{-3U}\biggl[i(\Dsl X)\gamma^{a}
+4Xe^{W-V}\gamma_5\gamma^{a}\biggr]\psi^\conj_a 
\nonumber\\
&+24e^{-3U}\biggl[i\gamma_5(\Dsl X)
-4Xe^{W-V}\biggr]\lambda^\conj +6e^{-3U}\biggl[i\gamma_5(\Dsl X)
-4Xe^{W-V}\biggr]\varphi^\conj\, .
\end{align}

Putting the previous results together, we find that the set of equations for the $\lambda,\varphi$ and $\psi_{a}$ modes is given by 
\begin{align}
 {\cal L}_{gr}^{a} +\frac{1}{4} {\cal R}^{a}_{gr} &=0\label{eom original 1}\\
{\cal L}_{f} +\frac{1}{4} {\cal R}_{f} &=0\\
{\cal L}_{b} +\frac{1}{4} {\cal R}_{b} &=0\, .\label{eom original 3}
\end{align}

\noindent These equations can be greatly simplified by a suitable field redefinition which we perform below. For convenience, the resulting equations are written out in full in Appendix \ref{Appendix:eom}.

\subsection{Field redefinitions and diagonalization}
We now look for a set of fields that produce diagonal kinetic terms for the various modes. The derivative terms in the equations above can be obtained from a Lagrangian density  (with respect to the 4-$d$ Einstein measure $d^4x\sqrt{|g|}$) of the form\footnote{We leave the overall normalization of the Lagrangian unfixed. We note that, as usual, the kinetic terms are real up to a total derivative. In the context of holography, the boundary terms are crucial as they determine the on-shell action. These should be determined separately when necessary.}
\begin{align}
\mathcal{L}_{kin} ={}&  e^W\left[
\bar{\psi}_a\gamma^{abc}D_{b}\psi_{c}+\left(\bar{\varphi} + 6\bar{\lambda}\right)\gamma_{5}\gamma^{ab}D_{a}\psi_b   + \bar{\psi}_a \gamma_{5}\gamma^{ab}D_{b}\left(6\lambda + \varphi \right)\right.\nonumber\\
&\left.-6\bar{\varphi}\Dsl\lambda-6\bar{\lambda}\Dsl\left(5\lambda +\varphi\right)\right].
\end{align}
We can rewrite these terms in diagonal form by means of the following field redefinitions:
\begin{align}
\zeta_{a} &= e^{W/2}\left[\psi_{a} - \frac{1}{2}\gamma_{5}\gamma_{a}\left( \varphi+6\lambda\right)\right],\label{definition zeta}\\
\eta &= e^{W/2}(\varphi + 2\lambda),\\
\xi &= 6e^{W/2}\lambda\, ,\label{definition xi}
\end{align}
so that
\begin{align}\label{kinetic lagrangian}
\mathcal{L}_{kin} 
={}& \bar{\zeta}_a\gamma^{abc}D_{b}\zeta_{c}+\frac{3}{2}\bar{\eta}\Dsl\eta +  \frac{1}{2}\bar{\xi}\Dsl\xi-\frac{1}{2}\left[\bar{\zeta}_a\gamma^{abc}(\partial_{b}W)\zeta_{c}+\frac{3}{2}\bar{\eta}(\pasl W)\eta +  \frac{1}{2}\bar{\xi}(\pasl W)\xi\right].
\end{align}
The interaction terms are produced by the action of the derivatives on the warping factors involved in the field redefinitions, and they will cancel against similar terms in the interaction Lagrangian. In section \ref{section:N=2}, we will interpret the fields $\zeta_{a},\eta,\xi$ in terms of the multiplet content appropriate to the underlying supersymmetry of the $d=4$ theory. Finally, it is worth noting that given our conventions for charge conjugation (see section \ref{Appendix:ChargeConjugation}), the redefinition \eqref{definition zeta} implies that the corresponding charge conjugate field is given by
\begin{equation}
\zeta_a^{\conj}= e^{W/2}\left[\psi_a^{\conj}+\frac12\gamma_5 \gamma_a\left(\varphi^{\conj}+6\lambda^{\conj}\right)\right].
\end{equation}

\subsection{Effective $d=4$ action}

By taking appropriate linear combinations of \eqref{eom original 1}-\eqref{eom original 3}
one can obtain the equations of motion for the diagonal fermion fields \eqref{definition
zeta}-\eqref{definition xi}. The resulting equations are written explicitly in Appendix
\ref{Appendix:eom}, and can be obtained from the following $d=4$ action
functional:\footnote{In writing the action below, we have performed a chiral rotation of
the form $\psi \mapsto e^{i\pi \gamma_{5}/4}\psi$ in all three fermion fields. This
transformation introduces a factor of $i\gamma_{5}$ in all bilinears of the form
$\bar{\psi}\gamma_{a_{1}}\gamma_{a_{2}}\ldots \gamma_{a_{2k}}\psi$ and
$\bar{\psi}\gamma_{a_{1}}\gamma_{a_{2}}\ldots \gamma_{a_{2k}}\psi^{\conj}$, while leaving
the rest (e.g. kinetic terms) invariant. This rotation has the virtue of producing
standard Dirac mass terms in the truncations we review in section
\ref{section:Examples}.
}
\begin{equation}\label{effective action}
S_{F} = K\int d^{4}x\sqrt{-g}\left[\bar{\zeta}_a\gamma^{abc}D_{b}\zeta_{c}+\frac{3}{2}\bar{\eta}\Dsl\,\eta +  \frac{1}{2}\bar{\xi}\Dsl\,\xi + {\cal L}^{int}_{\bar{\psi}\psi} + \frac{1}{2}\left({\cal L}^{int}_{\bar{\psi}\psi^\conj} +  \mbox{c.c.} \right)\right],
\end{equation}
where $K$ is a normalization constant, `` + c.c." denotes the complex conjugate (or,
equivalently, the charge conjugate) of ${\cal L}^{int}_{\bar{\psi}\psi^\conj}$, and the
interaction pieces ${\cal L}^{int}_{\bar{\psi}\psi}$ and ${\cal
L}^{int}_{\bar{\psi}\psi^\conj}$ are defined as

\begin{align}\label{final fermion Lagrangian}
{\cal L}^{int}_{\bar{\psi}\psi}={}&
+\frac34 i (\pa_b h)e^{-2U-V}\bar{\zeta}_{a}\gamma_5\gamma^{abc}\zeta_c
+\frac{3}{8}ie^{-2U-V}\bar{\eta}\gamma_{5}(\pasl h)\eta
-\frac38ie^{-2U-V}\bar\xi\gamma_5(\pasl h)\xi
\nonumber\\
&+\frac14e^{-2W-V}{H_3}^{abc}\bar{\zeta}_{a}\gamma_{5}\gamma_b\zeta_c
-\frac{3}{8} e^{-2W-V}\bar{\eta}\gamma_{5}\Hsl_{3} \eta
+\frac{3}{8} e^{-2W-V}\bar{\xi}\gamma_{5}\Hsl_{3} \xi
\nonumber\\
&
-\frac{i}{4}\bar{\zeta}_{a}\biggl[ 6\left(\pasl U \right)+e^{-2W-V}\gamma_{5}\Hsl_{3}\biggr]\gamma^a\xi
+ \frac{i}{4}\bar{\xi}\gamma^{a}\biggl[6\left(\pasl U \right)-e^{-2W-V}\gamma_{5}\Hsl_{3}\biggr]\zeta_a
\nonumber\\
 &
-\frac{3}{4}e^{-2U-V}\left[\bar{\zeta}_{a}\gamma_{5}(\pasl T)\gamma^a\eta -
\bar{\eta}\gamma_{5}\gamma^{a}(\pasl T^{\dagger}) \zeta_a\right]
\nonumber\\
&
+\frac{i}{4}\bar{\zeta}_{a}\biggl[-
e^{V-W}\left(F+ i\gamma_{5}*F\right)^{ac}
+3ie^{-W-2U}\gamma_{5}(H_2+i\gamma_5 *H_2)^{ac}\biggr]\zeta_c
\nonumber\\
&
+\frac{3i}{4}e^{V-W}\bar{\eta}\left(\Fsl - i\gamma_{5}e^{-V-2U}\Hsl_{2}\right)\eta 
-\frac{i}{8}e^{V-W}\bar{\xi}\left(\Fsl + 3i\gamma_{5}e^{-V-2U}\Hsl_{2}\right)\xi
\nonumber\\
 &
+\frac{3}{8}e^{V-W}\biggl[\bar{\zeta}_{a}\left(\Fsl-i\gamma_{5}e^{-V-2U}\Hsl_{2}\right)\gamma^a\eta
+\bar{\eta}\gamma^{a}\left(\Fsl-i\gamma_{5}e^{-V-2U}\Hsl_{2}\right)\zeta_a \biggr]
\nonumber\\
&
 -3ie^{W-4U}\bar{\zeta}_{a}\gamma_{5}T^{\dagger}\gamma^{ac}\zeta_c
 +3ie^{W-4U}\bar\eta \gamma_{5}T^\dagger\eta
 +\frac32e^{W-4U}\left(\bar\zeta_a\gamma^a\gamma_5T\eta+\bar\eta T\gamma_5\gamma^a\zeta_a\right)
\nonumber\\
&
-\frac{9i}{2}e^{W-4U}\bar\xi\gamma_{5} T\xi
-3ie^{W-4U}(\bar\eta\gamma_{5} T\xi+\bar\xi\gamma_{5} T\eta)
 +3e^{W-4U}\left(\bar\zeta_a\gamma^a\gamma_5T\xi+\bar\xi T\gamma_5\gamma^a\zeta_a\right)
\nonumber\\
&
+\frac{1}{4}i\left(\tilde{f}  -8e^{W-V}\right)\left(
i\bar{\zeta}_{a}\gamma^{ac}\zeta_c
-3i\bar\eta\eta
+\frac32\bar\zeta_a\gamma^a\eta+\frac32\bar\eta\gamma^a\zeta_a
\right)
\nonumber\\
&
+\frac{1}{8}\left(3\tilde{f}+8e^{W-V} \right)\bar{\xi}\xi
 +\frac{3}{4}\tilde{f}\left(\bar{\eta}\xi +\bar\xi\eta \right)
 +\frac14i\tilde f\left(\bar\xi\gamma^a\zeta_a+\bar\zeta_a\gamma^a\xi\right)
\end{align}
and\footnote{Note that some of the terms written below are actually equal, but we have left them this way to make the $N=2$ structure of covariant derivatives more manifest. See the next section for details.}
\begin{align}\label{final fermion Lagrangian 2}
{\cal L}^{int}_{\bar{\psi}\psi^\conj} ={}
&
e^{-3U}\Biggl\{
-\frac{i}{2}(D_bX)\bar{\zeta}_{a}\gamma_{5}\gamma^{abc}\zeta^\conj_c
 -\frac{3i}{4}\bar{\eta}\gamma_{5} (\Dsl X)\eta^{\conj}
- \frac{1}{4}\bar{\zeta}_{a}\gamma_{5}(\Dsl X)\gamma^a\xi^{\conj}
+\frac{1}{4}\bar{\xi}\gamma^{a}\gamma_{5}(\Dsl X)\zeta^\conj_a
\Biggr\}
\nonumber\\
&
+Xe^{W-V-3U}\Biggl\{
2i\bar{\zeta}_{a}\gamma_{5}\gamma^{ac}\zeta^\conj_c
 -6i\bar{\eta}\gamma_{5}\eta^{\conj}
- \bar{\zeta}_{a}\gamma_{5} \gamma^a\xi^{\conj}
+\bar{\xi}\gamma^{a}\gamma_{5}\zeta^\conj_a
\nonumber\\
&
-3\Bigl[\bar{\zeta}_{a}\left( \gamma_{5}\gamma^{a}\right)\eta^{\conj} +\bar{\eta}\left(\gamma_{5}\gamma^{a}
\right)\zeta^\conj_a  +i\bar{\eta}\gamma_{5}\xi^\conj 
 +i\bar{\xi}\gamma_{5}\eta^{\conj}\Bigr]\Biggr\},
\end{align}

\noindent where we have introduced the shorthand
\begin{equation}
\tilde{f} \equiv fe^{-3W}  +6e^{W+V-2U}\, , \qquad T\equiv h - i\gamma_{5}e^{V+2U}\, .
\end{equation}

\noindent We recall that all the fermions have charge $\pm 2$ with respect to the graviphoton, so that $D_{a} = \nabla_{a} - 2iA_{a}$ when acting on $\zeta, \eta, \xi$, while the complex scalar $X$ has charge $-4$, i.e. $DX = dX - 4iAX$. It is worth noting that the action \eqref{effective action} is manifestly real (up to total derivatives), and that it can  also be obtained by directly reducing the action of $D=11$ supergravity to the $SU(3)$ singlet sector. In particular, this procedure fixes the normalization constant $K$ in terms of the volume of the KE base $Y$, the length of the fiber parameterized by $\chi$, the normalization of the internal spinors $\varepsilon_{\pm}$, and the eleven-dimensional gravitational constant.

\section{$N=2$ supersymmetry}\label{section:N=2}

To interpret this action further, we consider how the fields fit into supermultiplets of gauged $N=2$ supergravity in four dimensions, ignoring the possibility of supersymmetry enhancement for special compactifications. Using the same techniques as above, we can reduce the 11-$d$ supersymmetry variations of the fermionic fields.\footnote{In what follows we keep only the terms linear in fermions.} These take the form
\beq\label{gravitinovariation11}
\delta\Psi_A=\hat D_A\Theta+\frac{1}{12}\frac{1}{4!}({\Gamma_A}^{BCDE}-8\delta_A^B\Gamma^{CDE})\Theta F_{BCDE}\, .
\eeq
We are interested only in the Grassmann parameters that are $SU(3)$ invariant, and it proves convenient to then write 
\beq
\Theta=e^{W/2}\theta\otimes\varepsilon_+e^{2i\chi}+e^{W/2}\theta^\conj\otimes\varepsilon_-e^{-2i\chi}\, .
\eeq
Here, $\theta$ is a 4-$d$ Dirac spinor.
By making appropriate projections on (\ref{gravitinovariation11}) to terms of definite charge, one obtains the variations of the fields $\varphi,\lambda,\psi_a$. 
Performing then the change of variables \eqref{definition zeta}-\eqref{definition xi}, we arrive at the variations
\beqn
\delta\eta&=&
-\frac14e^{V-W}\left(\Fsl-ie^{-2U-V}\gamma_5\Hsl_2\right)\theta
+\frac{i}{2}e^{-2U-V} (\pasl T)\theta
\nonumber\\
&&
-e^{W-4U}T\gamma_5\theta
-\frac14i\left(\tilde f-8e^{W-V}\right)\theta
-2e^{W-3U-V}X\gamma_5\theta^\conj\label{deltaeta}
\\
\delta\xi&=&
 3\gamma_5(\pasl U)\theta 
 -\frac{1}{2}e^{6U}\Hsl_3\theta
-\frac{1}{2}e^{-3U}i(\Dsl X)\theta^\conj 
\nonumber\\
&& 
-\frac{1}{2}i\tilde f\theta
+6e^{W-4U}T^\dagger\gamma_5\theta
-2Xe^{W-V-3U}\gamma_5\theta^\conj \label{deltaxi}
\\
\delta\zeta_a&=&
\left( D_a-\frac34i(\pa_a h)e^{-2U-V}\gamma_5
+\frac18e^{V-W}\gamma_5\left(\Fsl-3ie^{-V-2U}\gamma_5\Hsl_2\right)\gamma_a\right)\theta
\nonumber\\
&&+\left(
\frac18i\left(\tilde f-8e^{W-V}\right)\gamma_5
+\frac32Te^{W-4U}
\right)\gamma_a\theta
+\frac{1}{8}e^{-2W-V}\gamma_5\left[\gamma_a,\Hsl_3\right]\theta
\nonumber\\
&&
-Xe^{W-3U-V}\gamma_a\theta^\conj
+\frac12e^{-3U}\gamma_5(iD_aX)\theta^\conj\, .
\label{deltazeta}
\eeqn
%

Now, according to \cite{Gauntlett:2009zw}, there is a single vector multiplet that contains the scalar $\tau=h+ie^{V+2U}$ (in this notation, $T=\tau P_- +\bar\tau P_+$, where $P_\pm=\frac12(1\pm\gamma_5)$), and there is universal hypermultiplet containing $\rho=4e^{6U}$, the pseudoscalar dual to $H_3$ and $X$. The gravity multiplet contains the gravitino $\zeta_a$ while the vector multiplet and hypermultiplet each contains a Dirac spinor. Examining then the first lines of the variations (\ref{deltaeta}) and (\ref{deltaxi}) written above which contain derivatives of bosonic fields, we can identify the gauginos with $\eta$ and the hyperinos with $\xi$. 

In the $N=2$ literature, one usually finds things written in terms of Weyl spinors. For a generic spinor $\Psi$, we could write
\beq
\Psi_1=P_+\Psi,\ \ \ \ \ \Psi_2=P_+\Psi^\conj
\eeq
and we then have  $\Psi_2^\conj=P_-\Psi$ and  $\Psi_1^\conj=P_-\Psi^\conj$. To be specific, let us consider  the gaugino variation. It is convenient to first write the charge conjugate equation
\beqn
\delta\eta^\conj&=&
-\frac14e^{V-W}\left(\Fsl-ie^{-2U-V}\gamma_5\Hsl_2\right)\theta^\conj
-\frac{i}{2}e^{-2U-V} (\pasl T)\theta^\conj
\nonumber\\
&&
+e^{W-4U}T\gamma_5\theta^\conj
+\frac14i\left(\tilde f-8e^{W-V}\right)\theta^\conj
+2e^{W-3U-V}X^*\gamma_5\theta
\eeqn
and doing the chiral projection, we then obtain
\beqn
\delta\eta_1&=&
+\frac{i}{2}e^{-2U-V} (\pasl \tau)\theta_1^\conj
-\frac14e^{V-W}\left(\Fsl-ie^{-2U-V}\Hsl_2\right)\theta_1
\nonumber\\
&&
-e^{W-4U}\bar\tau\theta_1
-\frac14i\left(\tilde f-8e^{W-V}\right)\theta_1
-2e^{W-3U-V}X\theta_2
\\
\delta\eta_2&=&
-\frac{i}{2}e^{-2U-V} (\pasl \tau)\theta_2^\conj
-\frac14e^{V-W}\left(\Fsl-ie^{-2U-V}\Hsl_2\right)\theta_2
\nonumber\\
&&
+e^{W-4U}\bar\tau\theta_2
+\frac14i\left(\tilde f-8e^{W-V}\right)\theta_2
+2e^{W-3U-V}X^*\theta_1\, .
\eeqn
With a minor change of notation, these expressions can be understood as those that are obtained from working out this specific case of Ref. \cite{Andrianopoli:1996cm}. (Details of the bosonic sector of this have also recently appeared in Ref. \cite{Bobev:2010ib}). Indeed, we have worked through the details of deriving the 4-$d$ action using the results of \cite{Andrianopoli:1996cm}; we will not show this calculation in full here, but just point out the geometric features. The field content is usually presented after dualizing $H_2$ and $H_3$ \cite{Gauntlett:2009bh}\footnote{It's convenient to note that these imply
\beqn
\Hsl_2&=&\frac{ h+T}{|h+\tau|^2}(\tilde \Hsl_2+h^2\Fsl)\\
ie^{6U}\gamma_5\Hsl_3&=&\frac{1}{\rho}\left[ \Dsl \sigma+\Jsl_X\right]
\eeqn
}
\beqn
    H^{(2)}&=&\frac{1}{4h^2+e^{4U+2V}}\left(2h(\tilde{H}^{(2)}+h^2F^{(2)})-e^{2U+V}*(\tilde{H}^{(2)}+h^2F^{(2)})\right)\\
    H^{(3)}&=&-\frac14e^{-12U}*\left[D\sigma+J_X\right]
\eeqn
where $Da=da+6(\tilde B_1-\epsilon A_1)$, $\tilde H_2=d\tilde B_1$, $J_X=i(X^*DX-DX^* X)$, $\rho=4e^{6U}$ and $\sigma =4a$. The hypermultiplet contains the scalars $\{X,\sigma,\rho\}$, while the vector multiplet contains $\tau=h+ie^{V+2U}$. The scalars of the hypermultiplet coordinatize a quaternionic space ${\cal HM}\simeq SO(4,1)/SO(4)$ with metric
\beq
ds_{\cal H}^2=\frac{1}{\rho^2}d\rho^2+\frac{1}{4\rho^2}\left[d\sigma-i\left(X dX^*-X^*dX\right)\right]^2+\frac{1}{\rho^2}dX dX^*\, .
\eeq
The vector multiplet scalars coordinatize a special K\"ahler manifold ${\cal SM}$ with K\"ahler potential
\begin{equation}
    K_V=-\log \frac{i(\tau-\bar{\tau})^3}{2}\, .
\end{equation}
On ${\cal SM}$ there is a line bundle ${\cal L}$ with $c_1({\cal L})=\frac{i}{2\pi}\bar\pa\pa K_V=\frac{3i}{8\pi}\frac{1}{(Im\tau)^2}$.  Each of the fermions is a section of ${\cal L}^{1/2}$, with Hermitian connection $\theta=\pa K_V $. In the local coordinates $\tau,\bar\tau$, we have $\theta=-\frac{3}{2iIm\tau}d\tau$. Associated naturally to the line bundle is a $U(1)$ bundle with connection ${\cal Q}=Im\theta=\frac{3}{2} \frac{ dRe\tau}{Im\tau}$. Given $\tau=h+ie^{V+2U}$, this gives ${\cal Q}=\frac32 e^{-V-2U}dh$. 
The gaugino is also a section of $T{\cal SM}$; the Levi-Civita connection on ${\cal SM}$ is $\Gamma\equiv{\Gamma^\tau}_\tau=\frac{i}{Im\tau}d\tau=ie^{-V-2U}dh-d(V+2U)$. 

Because of the quaternionic structure, ${\cal HM}$ possesses three complex structures ${\cal J}^\alpha:T{\cal HM}\to T{\cal HM}$ that satisfy the quaternion algebra ${\cal J}^\alpha {\cal J}^\beta=-\delta^{\alpha\beta}1+\epsilon^{\alpha\beta\gamma}{\cal J}^\gamma$. Correspondingly, there is a triplet of K\"ahler forms $K_H^\alpha$, which we regard as $SU(2)$ Lie algebra valued. Required by $N=2$ supersymmetry, there is a principal $SU(2)$-bundle ${\cal SU}$ over ${\cal HM}$ with connection such that the hyper-K\"ahler form is covariantly closed; the curvature of the principal bundle is proportional to the hyper-K\"ahler form. It follows that the Levi-Civita connection of ${\cal HM}$ has holonomy contained in $SU(2)\otimes Sp(2,\mathbb{R})$. The fermions are sections of these bundles as follows:
\begin{itemize}
  \item gravitino: $\mathcal{L}^{1/2}\times\mathcal{SU}$
  \item gaugino: $\mathcal{L}^{1/2}\times\mathcal{TSM}\times\mathcal{SU}$
  \item hyperino: $\mathcal{L}^{1/2}\times\mathcal{THM}\times\mathcal{SU}^{-1}$
\end{itemize}
In the last line, one means that the hyperino is a section of the vector bundle obtained by deleting the $SU(2)$ part of the holonomy group on ${\cal HM}$.

The connections on ${\cal SU}$ and $T{\cal HM}\times {\cal SU}^{-1}$ are evaluated in terms of the hypermultiplet scalars, and one finds the following results, following a translation into Dirac notation. The gravitino covariant derivative reads
\begin{equation}
{\cal D}_{b} \zeta_{c}= D_{b}\zeta_{c} - \frac{3i}{4}e^{-2U-V}(\partial_{b}h)\gamma_{5}\zeta_{c} -\frac{i}{4}e^{6U}(*H_{3})_{b}\zeta_{c}+\frac{i}{2}e^{-3U}(D_{b}X)\gamma_{5}\zeta_{c}^{\conj}\, ,
\end{equation}
which leads to
\begin{align}
\gamma^{abc}{\cal D}_{b} \zeta_{c} ={}& \gamma^{abc}D_{b}\zeta_{c} +\frac{3i}{4}e^{-2U-V}(\partial_{b}h)\gamma_{5}\gamma^{abc}\zeta_{c} +\frac{1}{4}e^{6U}H_{3}^{abc}\gamma_{5}\gamma_{b}\zeta_{c}
\nonumber\\
&-\frac{i}{2}e^{-3U}(D_{b}X)\gamma_{5}\gamma^{abc}\zeta_{c}^{\conj}\, .
\end{align}
The gaugino covariant derivative is
\begin{equation}
{\cal D}_{a}\eta =D_{a}\eta -\frac{i}{4}e^{-(2U+V)}(\partial_{a}h)\gamma_5\eta -\frac{i}{4}e^{6U}(*H_{3})_{a}\eta+\frac{i}{2}e^{-3U}(D_{a}X)\gamma_5\eta^\conj\, ,
\end{equation}
giving
\begin{equation}
\cDsl\eta =\Dsl\,\eta +\frac{i}{4}e^{-(2U+V)}\gamma_5(\pasl h)\eta -\frac{1}{4}e^{6U}\gamma_{5}\Hsl_{3}\eta - \frac{i}{2}e^{-3U}\gamma_5(\Dsl X)\eta^\conj\, .
\end{equation}
Finally, the hyperino is a section of $T{\cal HM}\times {\cal SU}^{-1}$. The covariant derivative is then
\begin{equation}
{\cal D}_{a}\xi =D_{a}\xi +\frac{3i}{4}e^{-(2U+V)}(\partial_{a}h)\gamma_5\xi +\frac{3i}{4}e^{6U}(*H_{3})_{a}\xi\, .
\end{equation}
Equivalently,
\begin{equation}
\cDsl\xi =\Dsl\,\xi -\frac{3i}{4}e^{-(2U+V)}\gamma_5(\pasl h)\xi +\frac{3}{4}e^{6U}\gamma_{5}\Hsl_{3}\xi\, .
\end{equation}

We recognize the pieces of these covariant derivatives in the action given above. Indeed, the action takes the form
\begin{equation}
S_{kin} = K\int d^{4}x\sqrt{-g}\left[\bar{\zeta}_a\gamma^{abc}{\cal D}_{b}\zeta_{c}+\frac{3}{2}\bar{\eta}\cDsl\eta +  \frac{1}{2}\bar{\xi}\cDsl\xi+\cdots\right].
\end{equation}
In comparing to the first few lines of (\ref{final fermion Lagrangian}) and (\ref{final fermion Lagrangian 2}), one can see these covariant derivatives forming.
The remaining couplings to $F$ and $H_2$ and to the scalars can also be derived from the $N=2$ geometric structure, but we will not give further details here.

\section{Examples}\label{section:Examples}

In this section we compare the general effective four-dimensional action to various  holographic fermion systems that have been considered in the literature, and look for appropriate further (consistent) truncations  of the fermionic sector. We focus mainly on
two relevant further truncations, namely, the minimal gauged $N=2$ supergravity theory, and the model of \cite{Gauntlett:2009dn,Gauntlett:2009bh}, which provided an embedding of the holographic superconductor \cite{Hartnoll:2008vx,Hartnoll:2008kx} into M-theory.

\subsection{Minimal gauged supergravity}
As discussed in \cite{Gauntlett:2009zw}, a possible further truncation entails taking 
\begin{equation}
U=V =W=H_{3}= h =X = 0,\qquad f = 6\epsilon\, , \qquad   H_{2} = -\epsilon *F\quad (\mbox{i.e. } i\gamma_{5}\Hsl_{2} = \epsilon\Fsl)\, ,
\end{equation}
\noindent which sets all the massive fields to zero, leaving the $N=2$ gravity multiplet only. The corresponding equations for the bosonic fields can be derived from the Einstein-Maxwell action 
\begin{equation}
S_B =K_{B} \int d^{4}x\,\sqrt{-g}\left(R - F_{\mu\nu}F^{\mu\nu} + 24\right).
\end{equation}
The simplest fermionic content that one can consider is a charged massive bulk Dirac fermion minimally coupled to gravity and the gauge field (see for example \cite{Liu:2009dm}, \cite{Faulkner:2009wj}, \cite{Chen:2009pt}, \cite{Gubser:2009dt}, \cite{Gubser:2010dm}).

In our context, this truncation has an $AdS_{4}$ vacuum solution which uplifts to a supersymmetric $AdS_{4}\times SE_{7}$ solution in $D=11$. These solutions are thought of as being dual to three-dimensional SCFTs with $N = 2$ supersymmetry (in principle). In this truncation, we note that for $\epsilon=+1$, the variations (\ref{deltaeta}-\ref{deltaxi}) of $\eta$ and $\xi$ are both zero, and $\zeta_a$ decouples from $\eta,\xi$. Consequently, it is consistent to set $\eta=\xi=0$ (as we did for their superpartners) in this case,  and we then obtain the effective $d=4$ action \eqref{effective action} for the gravity supermultiplet 
\begin{equation}\label{effective action minimal SUGRA}
S = S_B+ K\int d^{4}x \sqrt{-g}\left[\bar{\zeta}_a\gamma^{abc} D_{b}\zeta_{c}  
-i\bar{\zeta}_{a} \bigl[\left(F +i\gamma_{5}*F\right)^{ac}+2i\gamma^{ac}\bigr]\zeta_c
 \right].
\end{equation}
We note that this gives the expected couplings between the gravitino and the graviphoton\footnote{One can use the identity $F^{bd}\gamma_{[b}\gamma^{ac}\gamma_{d]}= F_{bd}\gamma^{bdac}+2F^{ac}=i F_{bd}\gamma_5\epsilon^{bdac}+2F^{ac}$ to rewrite the coupling of the gravitino to the field-strength in the somewhat more familiar form $\sim F^{bd}\bar{\zeta}_{a}\gamma_{[b}\gamma^{ac}\gamma_{d]}\zeta_{c}\,$.} \cite{Freedman:1976aw},\cite{Fradkin:1976xz} (see \cite{Romans:1991nq} also).

If $\epsilon=-1$,  supersymmety is broken, and we wish to consider other truncations of the fermionic sector. It appears that there are no non-trivial consistent truncations in this case -- if we choose to set the gravitino to zero for example, its equation of motion gives a constraint on $\eta$ and $\xi$ that appears to have no non-trivial solutions.
To see this, we note the action contains the interaction terms (as usual neglecting 4-fermion couplings)
\begin{align}
{\cal L}^{int}_{\bar{\psi}\psi}={}&
 5\bar{\zeta}_{a}\gamma^{ac}\zeta_c
 -\frac92i\left(\bar\zeta_a\gamma^a\eta+\bar\eta\gamma^a\zeta_a\right)
 -3i\left(\bar\zeta_a\gamma^a\xi+\bar\xi \gamma^a\zeta_a\right)
\nonumber\\
 &
+\frac{i}{2}\bar{\zeta}_{a}\biggl[(F+ i\gamma_{5}*F)^{ac}\biggr]\zeta_c
+\frac{3}{4}\biggl[\bar{\zeta}_{a}\Fsl\gamma^a\eta
+\bar{\eta}\gamma^{a}\Fsl\zeta_a \biggr]
\nonumber\\
&
 -9\bar\eta\eta
-\frac72\bar\xi \xi
-3(\bar\eta \xi+\bar\xi \eta)
+\frac{3i}{2}\bar{\eta}\Fsl \eta 
+\frac{i}{4}\bar{\xi}\Fsl \xi\, .
\end{align}


\subsection{Fermions coupled to the holographic superconductor}

We now consider truncations appropriate to holographic superconductors. We note that the general model contains the charged boson $X$, of charge twice the charge of the fermion fields. This is one of the basic features of the model considered in \cite{Faulkner:2009am}, which studied charged fermions coupled to the holographic superconductor.
It is interesting to see how the couplings used there appear in the top-down model.

Refs. \cite{Gauntlett:2009dn,Gauntlett:2009bh} considered the following truncation of the bosonic sector
\begin{align}\label{the superconductor truncation}
h &= 0\, , \quad e^{6U} = 1 - \frac{1}{4}|X |^{2}\, , \quad V = -2U\,\, (=W), \quad H_{2} = *F\, , \nonumber\\
H_{3} &= \frac{i}{4}e^{-12U}*\left(X^* DX - XDX^{*}\right), \quad \epsilon = -1\, ,\quad  f = 6e^{-12U}\left(-1 + \frac{|X|^{2}}{3}\right),
\end{align}

\noindent where $DX = dX - 4iAX$ as before. As pointed out in \cite{Gauntlett:2009dn,Gauntlett:2009bh}, in order to set $h=0$ we need to impose $F\wedge F = 0$ by hand, and thus the truncation (even before considering the fermions) is not consistent. While this restriction allows for black hole solutions carrying electric or magnetic charge only, it excludes solutions of the dyonic type. This theory also has an $AdS_{4}$ vacuum solution (with $X=0$ and $f=-6$), which uplifts to a skew-whiffed $AdS_{4}\times SE_{7}$ solution in $D=11$. In general, these solutions do not preserve any supersymmetries (an exception being the case where $SE_{7} = S^{7}$). 

The $d=4$ effective action \eqref{effective action} for this truncation is given by
 \begin{equation}
S_{F} = K\int d^{4}x\sqrt{-g}\left[\bar{\zeta}_a\gamma^{abc} D_{b}\zeta_{c}+\frac{3}{2}\bar{\eta}\Dsl\, \eta +  \frac{1}{2}\bar{\xi}\Dsl\, \xi + {\cal L}^{int}_{\bar{\psi}\psi} + \frac{1}{2}\left({\cal L}^{int}_{\bar{\psi}\psi^\conj} +  \mbox{c.c.} \right)\right],
\end{equation}
\noindent where now
\begin{align}
e^{6U}{\cal L}^{int}_{\bar{\psi}\psi}={}&\frac{1}{2}\bar{\zeta}_{a}\Biggl[\left(1 - \frac{|X|^{2}}{4}\right)i\left(F +i\gamma_{5}*F\right)^{ac}
-2\left( |X|^{2}-5 \right)\gamma^{ac}
-\frac{1}{8}\left(X^{*}\overleftrightarrow{D_{b}}X\right)\gamma^{bac}
\Biggr]\zeta_c
\nonumber\\
&+\frac{3}{4}\bar{\eta}\Biggl[-4\left(3 - |X|^{2}\right) + \frac{1}{8} \left(X^{*}\overleftrightarrow{\Dsl}X\right) +2\left(1-\frac{|X|^{2}}{4}\right)i\Fsl\Biggr]\eta \nonumber\\
&+\frac{3}{8}\bar{\xi}\Biggl[-\frac{4}{3}\left(7 -|X|^{2}\right) +\frac{2}{3}\left(1 - \frac{|X|^{2}}{4}\right)i\Fsl
 -\frac{1}{4} \left(X^{*}\overleftrightarrow{\Dsl}X\right)
\Biggr]\xi
\nonumber\\
 &
+\frac{3}{4}\bar{\zeta}_{a}\biggl[2i\left(  |X|^{2}-3\right) +\left(1 - \frac{|X|^{2}}{4}\right)\Fsl
 \biggr]\gamma^a\eta +\frac{3}{4}\bar{\eta}\gamma^{a}\biggl[2i\left( |X|^{2}-3 \right) +\left(1 - \frac{|X|^{2}}{4}\right)\Fsl 
\biggr] \zeta_a
\nonumber\\
&+\frac{i}{2}\bar{\zeta}_{a}\left[ \left(|X|^{2}
-6\right) +\frac{1}{4} X^{*}(\Dsl X)\right]\gamma^a\xi+ \frac{i}{2}\bar{\xi}\gamma^{a}\left[\left(|X|^{2}-6\right)-\frac{1}{4}X(\Dsl X)^{*}
 \right]\zeta_a  \nonumber\\
 & 
 -\frac{3}{2}\bar{\eta}\left(2 - |X|^{2}\right)
 \xi 
 -\frac{3}{2}\bar{\xi}\left(2-|X|^{2}\right)\eta
\end{align}
\noindent and
\begin{align}
e^{3U}{\cal L}^{int}_{\bar{\psi}\psi^\conj} ={}&\frac{i}{2}\bar{\zeta}_{a}\gamma_{5}\left[-(D_bX)\gamma^{abc}+4X\gamma^{ac}
\right]\zeta^\conj_c -\frac{3i}{4}\bar{\eta}\gamma_{5}\left(\Dsl X+8X\right)\eta^{\conj}
\nonumber\\
&
- \frac{1}{4}\bar{\zeta}_{a}\gamma_{5}\left(\Dsl X+4X\right) \gamma^a\xi^{\conj}-\frac{1}{4}\bar{\xi}\gamma_{5}\gamma^{a}\left(\Dsl X +4X\right)\zeta^\conj_a
\nonumber\\
&-3X\biggl[\bar{\zeta}_{a}\left( \gamma_{5}\gamma^{a}\right)\eta^{\conj} +\bar{\eta}\left(\gamma_{5}\gamma^{a}
\right)\zeta^\conj_a +i\bar{\eta}\gamma_{5}\xi^\conj 
 +i\bar{\xi}\gamma_{5}\eta^{\conj}\biggr].
\end{align}

In order to compare to phenomenologically motivated models, such as  the holographic superconductor models, it is instructive to expand in powers of the complex scalar $X$, it being natural to organize the action by engineering dimension. Since 4-fermi couplings are dimension 6 or higher, we will here keep all terms up to and including dimension five. Doing so we obtain
\begin{align}
{\cal L}^{int}_{\bar{\psi}\psi}\simeq {}&\frac{1}{2}i\bar{\zeta}_{a}\biggl[\left(F +i\gamma_{5}*F\right)^{ac}
-10i\gamma^{ac}\biggr]\zeta_c
+\frac{3}{2}i\bar{\eta}\left(6i  +\Fsl\right)\eta \nonumber\\
&+\frac{1}{4}i\bar{\xi}\left(14i +\Fsl \right)\xi
+\frac{3}{4}\bar{\zeta}_{a}\left(-6i +\Fsl
 \right)\gamma^a\eta +\frac{3}{4}\bar{\eta}\gamma^{a}\left(-6i+\Fsl 
\right) \zeta_a
\nonumber\\
&-3\left(\bar{\eta}\xi +\bar{\xi}\eta +i\bar{\zeta}_{a}\gamma^a\xi +i\bar{\xi}\gamma^{a}\zeta_a \right)
\nonumber\\
&
-\frac14i|X|^{2} \left[
i\bar{\zeta}_{a}\gamma^{ac}\zeta_c
-\frac{3}{2}\left(\bar{\zeta}_{a}\gamma^a\eta+\bar{\eta}\gamma^{a}\zeta_a\right)
+\left(\bar{\zeta}_{a}\gamma^a\xi+\bar{\xi}\gamma^{a}\zeta_a \right)
\right]
\nonumber\\
 &
+\frac34|X|^{2} \left[
\bar{\eta} \eta 
-\frac12\bar{\xi}\xi
+\left(\bar{\eta}\xi+\bar{\xi}\eta\right)
\right],
\end{align}
\noindent and
\begin{align}
{\cal L}^{int}_{\bar{\psi}\psi^\conj} \simeq{}&\frac{1}{2}i\bar{\zeta}_{a}\gamma_5\left[-(D_bX)\gamma^{abc}+4X\gamma^{ac}
\right]\zeta^\conj_c -\frac{3}{4}i\bar{\eta}\gamma_5\left(\Dsl X+8X\right)\eta^{\conj}
\nonumber\\
&
- \frac{1}{4}\bar{\zeta}_{a}\gamma_5\left(\Dsl X+4X\right) \gamma^a\xi^{\conj}
-\frac{1}{4}\bar{\xi}\gamma_{5}\gamma^a\left(\Dsl X +4X\right)\zeta^\conj_a
\nonumber\\
&-3X\biggl(\bar{\zeta}_{a} \gamma_{5}\gamma^{a}\eta^{\conj} +\bar{\eta}\gamma_{5}\gamma^{a}\zeta^\conj_a 
+i\bar{\eta}\gamma_5\xi^\conj 
 +i\bar{\xi}\gamma_5\eta^{\conj}\biggr).
\end{align}
Note that we have the same basic couplings as in \cite{Faulkner:2009am}: we have Majorana couplings between the doubly-charged boson $X$ and spin-$1/2$ fermions. The model is significantly more complicated for several reasons. First, we have kept here several species of spin-$1/2$ fermions, and they are also coupled to the gravitino. An exploration of this model holographically, or a further truncation of the model, would be of interest. We also note that there are generic terms of the form $\bar\psi \gamma_5\Dsl X\psi^\conj$. These could also be of interest holographically; first in the presence of a boundary chemical potential for $A$, such a coupling looks similar to the other Majorana coupling near the boundary. But it also would presumably be the most important coupling in non-homogeneous boundary configurations (such as would correspond to spin-wave, nematic order, etc.). We also note that there are generically the ``Pauli terms", involving dipole couplings of the fermions to the gauge field strength, which could have important effects in electric or magnetic backgrounds.

It is clear that dropping all of the fermions is a consistent truncation, at least as consistent as the bosonic truncation.
It is also apparently possible to keep all of the fermions, although the $h$ equation of motion will now give a condition including terms non-linear in fermions. It would be interesting to find other truncations of the fermion content. For example, can one reduce, say, to a single species of charged fermion, including the elimination of the gravitino?. If such a truncation exists, it is non-trivial. 
\section{Conclusions}\label{section:Conclusions}
In this paper, we have explicitly worked out the form of the fermionic action obtained from a consistent truncation of 11-$d$ supergravity on warped Sasaki-Einstein 7-manifolds, which should be thought of as the total space of a $Spin^{c}$ bundle over a K\"ahler-Einstein base. The consistent truncation is obtained by restricting to $SU(3)$-invariant excitations. We have checked that the resulting theory is consistent with what is expected from $N=2$ gauged supergravity in four dimensions, in the case where there is a single vector multiplet and a single hypermultiplet.

This work is relevant to the recent literature on holographic duals of three-dimensional strongly-coupled field theories, particularly to those in which fermions play a central role in the dynamics, such as in superconductors. The theory does contain interesting couplings of the Majorana type, similar to those considered in the literature, as well as some new ones. We have briefly considered several further truncations that are closer to bottom-up models that have been discussed in the literature. Generally, we have found that it is difficult to find truncations of the fermionic sector. In particular, the gravitino is typically coupled to the other fermion fields. As a result, in holographic studies, we expect to see a spin-$3/2$ operator in the dual theory (the boundary supercurrents, in supersymmetric cases), and given appropriate asymptotic bosonic configurations, this operator would mix with other fermionic operators. We have not done an exhaustive job of studying this decoupling problem however, and it would be of interest to do so and to consider a variety of holographic applications.

\newpage
\centerline{\bf Acknowledgments}

It is a pleasure to thank Riccardo Argurio, Jim Liu, Phil Szepietowski and Diana Vaman for helpful conversations. J.I.J. and R.G.L. are thankful to the Michigan Center for Theoretical Physics (MCTP) for their hospitality during different stages of this project. We have been informed by J. Sonner of an ongoing collaboration with J. Gauntlett and D. Waldram that has some overlap with the work presented here. R.G.L. is supported by DOE grant FG02-91-ER40709. J.I.J. and A.T.F. are supported by  Fulbright-CONICYT fellowships, and I.B. by a University of Michigan Rackham Science Award. L.P.Z and I.B. are partially supported by DOE grant DE-FG02-95ER40899.

\appendix
\section{Conventions and useful formulae}\label{Appendix:Conventions}
In this Appendix we introduce the various conventions used in the body of the paper, and collect some useful results.

\subsection{Conventions for forms and Hodge duality}
We normalize all the (real) form fields according to
\begin{align}\label{rewriting rform in the wedge basis}
    \omega &= \omega_{a_{1}\ldots a_{p}}\,e^{a_{1}}\otimes e^{a_{2}}\cdots \otimes e^{a_{p}}\nonumber\\
           &= \frac{1}{p!}\omega_{a_1\ldots a_{p}} \,e^{a_{1}}\wedge \cdots \wedge e^{a_{p}}\, .
\end{align}

\noindent In $d$ spacetime dimensions, the Hodge dual acts on the basis of forms as
\begin{equation}
*(e^{a_1}\wedge \cdots \wedge e^{a_p}) = \frac{1}{(d-p)!}{\epsilon_{b_{1} \ldots b_{d-p}}}^{a_1\ldots a_p}\, e^{b_{1}}\wedge \cdots \wedge e^{b_{d-p}}\, ,
\end{equation}

\noindent where $\epsilon_{b_{1}\ldots b_{d-p}a_{1}\ldots a_{p}}$ are the components of the Levi-Civita \textit{tensor}. Equivalently, for the components of the Hodge dual $*\omega$ of a $p$-form $\omega$ we have
\begin{equation}\label{symbolicall meaning of the isomorphism via hodge}
	 (*\omega)_{a_{1}\ldots a_{d - p}} = \frac{1}{p!}{\epsilon_{a_{1}\ldots a_{d-p}}}^{b_{1}\ldots b_{p}}\omega_{b_{1}\ldots b_{p}}\, .
\end{equation}

\noindent In the $(3+1)$-dimensional external manifold $M$ we adopt the convention $\epsilon_{0123} = +1$ for the components of the Levi-Civita tensor in the orthonormal frame. 

\subsection{Elfbein and spin connection}
As discussed in section \ref{section:Ansatz}, the Kaluza-Klein metric ansatz of \cite{Gauntlett:2009zw} is given by
\begin{equation}
ds^2_{11}=e^{2W(x)}ds_{E}^2(M)+e^{2U(x)}ds^2(Y)+e^{2V(x)}\bigl(d\chi + {\cal A}(y) +A(x)\bigr)^2\, ,
\end{equation}

\noindent where $W(x) = -3U(x) - V(x)/2$ as in the body of the paper. We now introduce the eleven-dimensional orthonormal frame $\hat{e}^{M}$. Denoting by $a,b,\ldots$ the tangent indices to $M$, by $\alpha,\beta,\ldots$ the tangent indices to the KE base $Y$, and by $f$ the index associated with the U(1) fiber direction $\chi$, our choice of elfbein reads
\begin{align}
\hat e^a &= e^{W}e^a\\
\hat e^\alpha &= e^U e^\alpha\\
\hat e^f &= e^V \Bigl(d\chi + {\cal A}(y) +A(x)\Bigr),
\end{align}

\noindent where $e^{a}$ and $e^{\alpha}$ are orthonormal frames for $M$ and $Y$, respectively. The dual basis is then
\begin{align}
\hat e_a &= e^{-W}\bigl(e_a-A_a\pa_\chi\bigr) \\
\hat e_\alpha &= e^{-U}\bigl( e_\alpha-{\cal A}_\alpha\pa_\chi\bigr)\\
\hat e_f &= e^{-V}\pa_\chi \, .
\end{align}

\noindent Denoting by ${\omega^a}_b$ the spin connection associated with $ds^2(M)$ and by ${\omega^\alpha}_\beta$ the spin connection appropriate to $ds^2(Y)$, for the eleven-dimensional spin connection ${\hat{\omega}^{M}}_{~ \, N}$ we find
\begin{align}
{\hat\omega^\alpha}_{~a} &= e^{U-W}(\pa_a U)e^\alpha \\
\hat\omega^f_{~a} &= e^{V-W}\left[\frac12F_{ab}e^b+(\pa_aV)\bigl(d\chi + {\cal A}+A\bigr)\right] \\
\hat\omega^f_{~\alpha} &= e^{V-U}\frac12{\cal F}_{\alpha\beta}e^\beta \\
\hat\omega^a_{~b} &= \omega^a_{~b}-2\eta^{ac}\pa_{[c}W \eta_{b]d}e^{d}-\frac12 e^{2(V-W)}F^a_{~b}\bigl(d\chi+{\cal A}+A\bigr)\\
\hat\omega^\alpha_{~\beta} &= \omega^\alpha_{~\beta}-\frac12 e^{2(V-U)}{\cal F}^\alpha_{~\beta}\bigl(d\chi+{\cal A}+A\bigr),
\end{align}

\noindent where $\eta_{ab}$ is the flat metric in $(3+1)$ dimensions, $F \equiv dA$ and ${\cal F} \equiv d{\cal A} = 2J$, $J$ being the K\"ahler form on $Y$. 

\subsection{Fluxes}\label{Appendix:Conventions-Fluxes}
The ansatz \eqref{F4 ansatz} for the 4-form flux $\hat{F}_{4}$, reproduced here for convenience, is \cite{Gauntlett:2009zw} 
\begin{align}\label{F4 ansatz revisited}
\hat F_4={}&f \, \mbox{vol}_4+H_3\wedge (\eta + A)+H_2\wedge J+dh\wedge J\wedge (\eta + A)+2h J^2\nonumber\\
&+\left[X(\eta + A)\wedge\Omega-\frac{i}{4}\left(dX-4iAX\right)\wedge\Omega + \mbox{c.c.}\right].
\end{align}

\noindent We will often use a complex basis on $T^{*}Y$. If $y$ denote real coordinates on $Y$, we define $z^{1} \equiv \frac{1}{2}(y^{1} + iy^{2})$, $z^{\bar{1}} \equiv \frac{1}{2}\left(y^{1} - iy^{2}\right)$, and similarly for $z^{2},z^{\bar{2}},z^{3},z^{\bar{3}}$. With this normalization, the K\"ahler form $J$ and the holomorphic (3,0)-form $\Sigma$ are given by
\begin{align}
J &= 2i\sum_{\alpha = 1,2,3}e^{\alpha}\wedge e^{\bar{\alpha}}\, \\
\Sigma &= \frac{8}{3!}\epsilon_{\alpha\beta\gamma}\, e^\alpha\wedge e^\beta\wedge e^\gamma\, ,\label{defSigma}
\end{align}

\noindent where we have chosen $\epsilon_{123} = +1$. Similarly, the forms on the external manifold can be written
\begin{align}
\mbox{vol}_4 &=\frac{1}{4!}\epsilon_{abcd}\, e^a\wedge e^b\wedge e^c\wedge e^d \\
H_2 &=\frac{1}{2!}H_{2\,ab}\, e^a\wedge e^b \\
H_3 &=\frac{1}{3!}H_{3\,abc}\, e^a\wedge e^b\wedge e^c\, .
\end{align}
The components of $\hat{F}_{4}$ with respect to the eleven-dimensional frame $\hat{e}^{M}$ are then (in the real basis for $T^{*}Y$)
\begin{align}
\hat{F}_{abcf} &= e^{-3W-V}H_{3\, abc}\\
\hat{F}_{a\alpha\beta f}  &= e^{-W-2U-V}(\pa_a h) J_{\alpha\beta}\\
\hat{F}_{f\alpha\beta\gamma}  &= Xe^{-3U-V}\Omega_{\alpha\beta\gamma} + \mbox{c.c.}\\
\hat{F}_{abcd}  &= fe^{-4W}\epsilon_{abcd}\\
\hat{F}_{ab\alpha\beta} &= e^{-2W-2U}J_{\alpha\beta}H_{2\, ab}\\
\hat{F}_{\alpha\beta\gamma\delta} &= 4he^{-4U}(J_{\alpha\beta}J_{\gamma\delta}-J_{\alpha\gamma}J_{\beta\delta}+J_{\alpha\delta}J_{\beta\gamma})\\
\hat{F}_{a\alpha\beta\gamma} &= -\frac{i}{4}(D_aX)e^{-3U-W}\Omega_{\alpha\beta\gamma} + \mbox{c.c.}
\end{align}

\subsection{Clifford algebra}

 We choose the following basis for the $D=11$ Clifford algebra:
\begin{align}
 \Gamma^a &= \gamma^a\otimes \mathds{1}_{8}\\
 \Gamma^\alpha &= \gamma_5\otimes \gamma^\alpha\\
 \Gamma_f &=\gamma_5\otimes \gamma_7
\end{align}
 where the $\{\gamma^a\}$ are a basis for $C\ell(3,1)$ with $\gamma_5=i \gamma^0\gamma^1\gamma^2\gamma^3$ and the $\{\gamma^\alpha\}$ are a basis for $C\ell(6)$ with $\gamma_7=i\prod_\alpha\gamma^\alpha$. These dimensions are such that we can define Majorana spinors in each case. In $D=11$, we take $\Gamma^0$ to be anti-Hermitian and the rest Hermitian. This means that $\gamma^0$ is anti-Hermitian, while $\gamma^a (a\neq 0)$, $\gamma_5$, $\gamma_7$ and $\gamma^\alpha$ are Hermitian. We also have $\gamma_5^2=1$ and $\gamma_7^2=1$. In the standard basis, the $\{\gamma^a, \gamma_5\}$ are $4\times 4$ matrices while the $\{\gamma^\alpha, \gamma_7\}$ are $8\times 8$ matrices. It will also be convenient to define
\begin{align}
 \Gamma_7 &=\prod_\alpha \Gamma^\alpha = \mathds{1}_{4}\otimes \gamma_7\\
 \Gamma_5 &=\prod_a \Gamma^a = \gamma_5\otimes \mathds{1}_{8}\, .
\end{align}
 

\noindent Some useful identities involving the $C\ell(3,1)$ gamma matrices include
\begin{equation}
\epsilon_{abcd} = -i\gamma_{5}\gamma_{abcd}\, ,\quad \epsilon_{abcd}\gamma^{a} = i\gamma_{5}\gamma_{bcd}\, , \quad\epsilon_{abcd}\gamma^{cd}=2i\gamma_{5}\gamma_{ab} \, ,\quad  \epsilon_{abcd}\gamma^{bcd} = 6i\gamma_{5}\gamma_{a}\, .
\end{equation}

\subsection{Charge conjugation conventions}\label{Appendix:ChargeConjugation}
In $d=4$ dimensions with signature $(-,+,+,+)$ we can define unitary intertwiners $B_{4}$ and $C_{4}$ (the charge conjugation matrix), unique up to a phase, satisfying
\begin{align}\label{summary intertwiner B summ}
B_{4}\gamma_{a}B_{4}^{\dagger} &=  \gamma^{*}_{a}&
B_{4}^{T} &= B_{4}\\
\label{summary intertwiner B 2 summ} B_{4}\gamma_{5} B_{4}^{\dagger} &=
-\gamma_{5}^{*}&
B_{4}^{*}B_{4} &=\mathds{1}\, ,& 
\end{align}

\noindent and 
\begin{align}\label{summary intertwiner C summ}
    C_{4}\gamma_{a}C_{4}^{\dagger} &= -\gamma^{T}_{a}&  C_{4}^{T} &= -C_{4}\\ \label{summary intertwiner C 2 summ}
    C_{4}\gamma_{5} C_{4}^{\dagger} &=\gamma_{5}^{T} &
    C_{4} &= B_{4}^{T}\gamma_{0}=B_{4}\gamma_{0}\, . & 
\end{align}

\noindent If $\psi$ is any spinor, its charge conjugate $\psi^{\conj}$  is then defined  as
\begin{equation}
\psi^{\conj} = B_{4}^{-1}\psi^{*}=B_{4}^{\dagger}\psi^{*} = \gamma_{0}C_{4}^{\dagger}\psi^{*}\, .
\end{equation}

\noindent In (3+1) dimensions one can define Majorana spinors. By definition, a spinor $\psi$ is Majorana if  $\psi = \psi^{\conj}$. Notice that in (3+1) dimensions this condition relates opposite chirality spinors. Similarly, we can define the charge conjugates of a spinor $\Psi$ in (10+1) dimensions and a spinor $\eta$ in $7$ Euclidean dimensions as
\begin{alignat}{5}
\Psi^{\conj} &= B_{11}^{-1}\Psi^{*}\, ,&  &\phantom{3445}& &\mbox{where}& &\phantom{3445}& B_{11}\Gamma_{M} B_{11}^{-1} &=\Gamma_{M}^{*}\, ,\\
\eta^{\conj} &= B_{7}^{-1}\eta^{*}\, ,&  &\phantom{3445}& &\mbox{where}& &\phantom{3445}& B_{7}\gamma_{\alpha} B_{7}^{-1} &=-\gamma_{\alpha}^{*}\,.
\end{alignat}

\noindent Defining $\psi^{\conj}$ in the (3+1)-dimensional space $M$ by using the intertwiner $B_{4}$ defined above, (as opposed to using an intertwiner $B_{4-}$ satisfying $B_{4-}\gamma_{a}B_{4-}^{\dagger} =  -\gamma^{*}_{a}$ and $B_{4-}^{T} = -B_{4-}$), ensures that the charge conjugation operation acts uniformly in all the 11 directions, with
\begin{equation}
B_{11} = B_{4}\otimes B_{7}\, .
\end{equation}

\section{More on $SU(3)$ singlets}\label{Appendix:singlets}

The crucial feature of the truncations we are examining is that we retain only singlets under the structure group of the KE base. To further understand the structure in play in the reduction of the fermionic degrees of freedom, we consider the corresponding problem on gravitino {\it states}.

In the complex basis, the $\Gamma$ matrices act as raising and lowering operators on the states. The raising operators transform as a ${\bf 3}$ of $SU(3)$ and the lowering operators as a ${\bf \bar 3}$. Using complex notation, we write $\Gamma^{1}=\frac12\left[\Gamma^1+i\Gamma^2\right]$, etc. where the matrix on the left-hand side is understood to be defined in the complex basis and those on the right are in the real basis. We then see that $\Gamma^\alpha$ and $\Gamma^{\bar \alpha}$ satisfy Heisenberg algebras, and we can associate Fock spaces to each pair. Then, $P_1=\Gamma^1\Gamma^{\bar 1}
$ is a projector, and we are led to define the set of projection operators (we are using complex indices, so $\alpha=1,2,3$)
\beq
P_\alpha=\Gamma^\alpha\Gamma^{\bar \alpha},\ \ \ \ \bar P_\alpha=\Gamma^{\bar \alpha}\Gamma^\alpha\ \ \ \ 
\mbox{(no sum)}
\eeq

\noindent and ``charge" operators \footnote{Note that $\Gamma^1,\Gamma^2,Q_1$ can be identified as the generators $J_x,J_y,J_z$ of the spin-$1/2$ representation of an $SU(2)$ subgroup, and similarly for $\Gamma^{3},\Gamma^{4},Q_{2}$, etc.} 
\beq
Q_\alpha=\Gamma^{\alpha\bar \alpha}\ \ \mbox{(no sum)}
\eeq

\noindent Since a spinor can be thought of in the corresponding Fock space representation as $|\pm~\frac12,\pm\frac12,\pm\frac12\rangle$, with the $\pm\frac12$ being eigenvalues of $Q_\alpha$, the $SU(3)$ singlets are those spinors that satisfy
\begin{align}
Q_\alpha \varepsilon_\pm=\pm\frac12\varepsilon_\pm, \ \ \ \forall \,\, \alpha
\end{align}
The six other states are in non-trivial representations of $SU(3)$.
Note that $\Gamma_7=\prod_\alpha 2Q_\alpha$, so the positive (negative) chirality spinor has an even (odd) number of minus signs, and $\Gamma_7$ is the ``volume form" (the product of all the signs). The (c-)spinors are in the ${\bf 4}+{\bf\bar 4}$ of $\mbox{Spin}(6)\simeq SU(4)$, with the two conjugate representations corresponding to the two chiral spinors. We can now appreciate the significance of the operator $Q$ that we encountered in section \ref{section:Ansatz}: it is (up to normalization) the ``total charge operator" $Q=2\sum_\alpha 2Q_\alpha$. It is clear that it is the $SU(3)$ singlets that have maximum charge $Q=\pm 6$, where the sign is correlated with the chirality. The other spinor states are in ${\bf 3}$ and ${\bf\bar 3}$ and have $Q$-charges $\mp 2$. We then find that the ordinary spinor consists of $\{ |{\bf 1},6\rangle_+, \{ |{\bf 3},-2\rangle_+, \{ |{\bf \bar 3},2\rangle_-, \{ |{\bf 1},-6\rangle_-\}$, where the subscript on the ket indicates the $\gamma_7$-chirality. In the weight language, the $|{\bf 1},6\rangle_+$ corresponds to $|\frac12,\frac12,\frac12\rangle$ and the $|{\bf 1},-6\rangle_-$  corresponds to $|-\frac12,-\frac12,-\frac12\rangle$, and it is clear from the construction that they are related by charge conjugation. 

As described in the body of the paper, we focus on the $SU(3)$ singlet spinors $\varepsilon_\pm$, and consequently discard all but the internal spinors
\begin{equation}
\varepsilon(y,\chi)=\varepsilon_\pm(y) e^{\pm 2i\chi}=\varepsilon_\pm(y) e^{\pm 2i\chi}\, .
\end{equation}

\noindent Notice that $\varepsilon_\pm$ are not only $\gamma_7$-chiral, but they satisfy the projections
\beq
\bar P_\alpha\varepsilon_+=0,\ \ \ \ P_\alpha\varepsilon_-=0,\ \ \ \forall\alpha
\eeq

Finally, the gravitino states can be thought of as the spin-1/2 spinor tensored with $ |{\bf 3},4\rangle\oplus  |{\bf \bar 3},-4\rangle$ (i.e. the representations corresponding to the raising/lowering operators). Thus, the gravitino states transform as 
$\{  |{\bf 3},10\rangle, |{\bf 1},6\rangle, |{\bf 8},6\rangle, |{\bf \bar 3},2\rangle,  |{\bf 6},2\rangle, |{\bf 3},-2\rangle\}$ and their conjugates. This totals 48 states, which is the right counting.

\section{$d=4$ equations of motion}\label{Appendix:eom}
Here we explicitly collect the equations of motion for the diagonal fermion fields $\zeta_{a},\eta$ and $\xi$. To this end we define the following linear combinations
\begin{align}
\mathcal{L}_{\zeta}^{a} &\equiv e^{\frac{3W}{2}}\gamma_{5}\mathcal{L}_{gr}^{a}& \mathcal{R}_{\zeta}^{a} &\equiv e^{\frac{3W}{2}}\gamma_{5}\mathcal{R}_{gr}^{a}\\
\mathcal{L}_{\eta} &\equiv e^{\frac{3W}{2}}\left(\frac{2}{3}\gamma_{5}\mathcal{L}_{f} + \frac{1}{3}\gamma_{a}\mathcal{L}^{a}_{gr}\right)& \mathcal{R}_{\eta} &\equiv e^{\frac{3W}{2}}\left(\frac{2}{3}\gamma_{5}\mathcal{R}_{f} + \frac{1}{3}\gamma_{a}\mathcal{R}_{gr}^{a}\right) \\
\mathcal{L}_{\xi}&\equiv \frac{2}{3}e^{\frac{3W}{2}}\left(\frac{1}{2}\mathcal{L}_{b}-\gamma_{5}\mathcal{L}_{f} + \gamma_{a}\mathcal{L}^{a}_{gr}\right)& \mathcal{R}_{\xi}&\equiv \frac{2}{3}e^{\frac{3W}{2}}\left(\frac{1}{2}\mathcal{R}_{b}-\gamma_{5}\mathcal{R}_{f} + \gamma_{a}\mathcal{R}_{gr}^{a}\right),
\end{align}

\noindent where $\mathcal{L}_{f},\mathcal{L}_{gr}^{a},\mathcal{L}_{b}$ and $\mathcal{R}_{f},\mathcal{R}_{gr}^{a},\mathcal{R}_{b}$ are given in section \ref{section:4d eqs}. After performing the chiral rotation of the fermion fields described in section \ref{section:4d eqs}, the equations of motion then read 
\begin{align}\label{E_chi}
0={}& \mathcal{L}_{\zeta}^{a} +\frac{1}{4}\mathcal{R}_{\zeta}^{a}\\
={}& \gamma^{abc} D_{b}\zeta_{c}+\frac{1}{4}\Biggl[
-ie^{V-W}\left(F + i\gamma_{5}*F\right)^{ac} -12ie^{W-4U}\gamma_{5}(h + i\gamma_{5}e^{V+2U})\gamma^{ac}
\nonumber\\
&\hphantom{\gamma^{abc} D_{b}\zeta_{c}+\frac{1}{4}\biggl[}\,+3i(\pa_bh)e^{-2U-V}\gamma_{5}\gamma^{abc}
-3e^{-W-2U}\gamma_{5}\left(H_2+i\gamma_5*H_2\right)^{ac}
\nonumber\\
&\hphantom{\gamma^{abc} D_{b}\zeta_{c}+\frac{1}{4}\biggl[}\,
-\left(fe^{-3W}+6e^{W+V-2U}  -8e^{W-V}\right)\gamma^{ac}
+e^{-2W-V}{H_3}^{abc}\gamma_{5}\gamma_b
\Biggr]\zeta_c
\nonumber\\
 &
+\frac{3}{8}\Biggl[i\left( fe^{-3W}+6e^{W+V-2U}-8e^{W-V}\right)+ e^{V-W}\left(\Fsl-i\gamma_{5}e^{-V-2U}\Hsl_{2}\right)
\nonumber\\
&
 \hphantom{+\frac{3}{8}\biggl\{}\, - 4e^{W-4U}\gamma_{5}\left(h + i\gamma_{5}e^{V+2U}\right)-2e^{-2U-V}\gamma_{5} \pasl\left( h - i\gamma_{5}e^{V+2U}\right)\Biggr]\gamma^a\eta\nonumber\\
&+\frac{1}{4}\biggl[ i\left(fe^{-3W}+6e^{W+V-2U}\right)
-12e^{W-4U}\gamma_{5}\left(h + i\gamma_{5}e^{V+2U}\right) -6i\left(\pasl U \right)\nonumber\\
&\hphantom{+\frac{1}{4}\biggl[}\,-ie^{-2W-V}\gamma_{5}\Hsl_{3}\biggr]\gamma^a\xi +\frac{i}{2}\gamma_{5}\left[-(D_bX)e^{-3U}\gamma^{abc}+4Xe^{W-3U-V}\gamma^{ac}
\right]\zeta^\conj_c 
\nonumber\\
&
- \frac{1}{4}\gamma_{5}\biggl[e^{-3U}(\Dsl X)\gamma^a+4Xe^{W-3U-V}\gamma^{a}\biggr] \xi^{\conj} -3Xe^{W-3U-V}\gamma_{5}\gamma^{a}\eta^{\conj}\, ,
\end{align}

\begin{align}\label{E_eta}
0 ={}&\mathcal{L}_{\eta} + \frac{1}{4}\mathcal{R}_{\eta} \nonumber\\
={}& \Dsl\,\eta +\Biggl[ \frac{1}{2}\left(fe^{-3W} +6e^{W+V-2U}- 8e^{W-V}\right) - \frac{1}{4}e^{-2W-V}\gamma_{5}\Hsl_{3} + \frac{i}{4}e^{-2U-V}\gamma_{5}(\pasl h)  \nonumber\\
&\hphantom{ \Dsl\,\eta+\Biggl[}\,+\frac{i}{2}e^{V-W}\left(\Fsl - i\gamma_{5}e^{-V-2U}\Hsl_{2}\right)+2i e^{W-4U}\gamma_{5}\left(h + i\gamma_{5}e^{V+2U}\right)\Biggr]\eta \nonumber\\
&+\frac{1}{4}\biggl[i\left(fe^{-3W} + 6e^{W+V-2U} -8e^{W-V}\right)
+ 4 e^{W-4U}\gamma_{5}\left(h-i\gamma_{5}e^{V+2U}\right)\biggr]\gamma^{b}\zeta_b
\nonumber\\
&+\frac{1}{4}\gamma^{b}\biggl[e^{V-W}\left(\Fsl-i\gamma_{5}e^{-V-2U}\Hsl_{2}\right) 
-2e^{-2U-V}\gamma_{5}\pasl\left( h + i\gamma_{5}e^{V+2U}\right)\biggr] \zeta_b
\nonumber\\
& 
 +\frac{1}{2}\biggl[\left(fe^{-3W}  +6e^{W+V-2U}\right)-4i\gamma_{5}e^{W-4U}\left(h - i\gamma_{5}e^{V + 2U}\right)
 \biggr]\xi 
\nonumber\\
&  - \frac{i}{2}\gamma_{5}\left[e^{-3U}(\Dsl X)+8Xe^{W-3U-V}\right]\eta^{\conj}+\left(-2 ie^{W-3U-V}X
\right)\gamma_{5}\xi^\conj 
\nonumber\\ 
&+\left(-2Xe^{W-3U-V}\gamma_{5}\gamma^{c}
\right)\zeta^\conj_c\, ,
\end{align}

\noindent and
\begin{align}\label{E_xi}
0 ={}& \mathcal{L}_{\xi}+\frac{1}{4}\mathcal{R}_{\xi}\nonumber \\
={}& \Dsl\,\xi +\frac{3}{4}\Biggl[\frac{8}{3}e^{W-V}  +\left(fe^{-3W} + 6e^{W+V-2U}\right)-\frac{i}{3}e^{V-W}\left(\Fsl + 3i\gamma_{5}e^{-V-2U}\Hsl_{2}\right)
\nonumber\\
&
\hphantom{\frac{1}{6}\Dsl\,\xi +\frac{1}{8}\Biggl[}\, +e^{-2W-V}\gamma_{5}\Hsl_{3} 
-12ie^{W-4U}\gamma_{5}\left(h - i\gamma_{5}e^{V+2U}\right)
-ie^{-2U-V}\gamma_{5}(\pasl h)\Biggr]\xi
\nonumber\\
&+ \frac{1}{2}\Biggl[i\gamma_{5}\gamma^{a}e^{-2W-V}\Hsl_{3} + 6i\gamma^{a}\left(\pasl U \right)
 +i\gamma^{a}\left(fe^{-3W} + 6e^{W+V-2U}\right) \nonumber\\
&
\hphantom{+ \frac{1}{12}\Biggl[}\,
+ 12e^{W-4U}\gamma_{5}\left(h - i\gamma_{5}e^{V+2U}\right)\gamma^{a}\Biggr]\zeta_a
\nonumber\\
&
 +\frac{3}{2}\biggl[\left(fe^{-3W}+6e^{W+V-2U}\right)
-4ie^{W-4U}\gamma_{5}\left(h-i\gamma_{5}e^{V+2U}\right)
\biggr]\eta\nonumber\\
& -\frac{e^{-3U}}{2}\gamma_{5}\gamma^{a}\biggl[(\Dsl X) +4Xe^{W-V}\biggr]\zeta^\conj_a -6iXe^{W-3U-V}\gamma_{5}\eta^{\conj} \, .
\end{align}

\noindent We recall that all the fermions have charge $\pm 2$ with respect to the graviphoton, so that $D_{a} = \nabla_{a} - 2iA_{a}$ when acting on $\zeta, \eta, \xi$, while the complex scalar $X$ has charge $-4$, i.e. $DX = dX - 4iAX$. Naturally, the equations of motion for the charge conjugate fields $\zeta^\conj_{a},\eta^\conj,\xi^\conj$ can be obtained by taking the complex conjugate of the equations above and using the rules given in section \ref{Appendix:ChargeConjugation}. Alternatively, the above equations can be obtained directly by taking functional derivatives of the effective action \eqref{effective action}.

\newpage
\providecommand{\href}[2]{#2}\begingroup\raggedright\endgroup

\providecommand{\href}[2]{#2}\begingroup\raggedright\begin{thebibliography}{10}

\bibitem{Maldacena:1997re}
J.~M. Maldacena, ``{The large N limit of superconformal field theories and
  supergravity},'' {\em Adv. Theor. Math. Phys.} {\bf 2} (1998)  231--252,
\href{http://arxiv.org/abs/hep-th/9711200}{{\tt arXiv:hep-th/9711200}}.

\bibitem{Gubser:1998bc}
S.~S. Gubser, I.~R. Klebanov, and A.~M. Polyakov, ``{Gauge theory correlators
  from non-critical string theory},''
  \href{http://dx.doi.org/10.1016/S0370-2693(98)00377-3}{{\em Phys. Lett.} {\bf
  B428} (1998)  105--114},
\href{http://arxiv.org/abs/hep-th/9802109}{{\tt arXiv:hep-th/9802109}}.

\bibitem{Witten:1998qj}
E.~Witten, ``{Anti-de Sitter space and holography},'' {\em Adv. Theor. Math.
  Phys.} {\bf 2} (1998)  253--291,
\href{http://arxiv.org/abs/hep-th/9802150}{{\tt arXiv:hep-th/9802150}}.

\bibitem{Aharony:1999ti}
O.~Aharony, S.~S. Gubser, J.~M. Maldacena, H.~Ooguri, and Y.~Oz, ``{Large N
  field theories, string theory and gravity},''
  \href{http://dx.doi.org/10.1016/S0370-1573(99)00083-6}{{\em Phys. Rept.} {\bf
  323} (2000)  183--386},
\href{http://arxiv.org/abs/hep-th/9905111}{{\tt arXiv:hep-th/9905111}}.

\bibitem{Klebanov2000}
I.~R. Klebanov and M.~J. Strassler, ``{Supergravity and a confining gauge
  theory: Duality cascades and chiSB-resolution of naked singularities},'' {\em
  JHEP} {\bf 08} (2000)  052,
\href{http://arxiv.org/abs/hep-th/0007191}{{\tt arXiv:hep-th/0007191}}.

\bibitem{Maldacena2001}
J.~M. Maldacena and C.~Nunez, ``{Towards the large N limit of pure N = 1 super
  Yang Mills},'' \href{http://dx.doi.org/10.1103/PhysRevLett.86.588}{{\em Phys.
  Rev. Lett.} {\bf 86} (2001)  588--591},
\href{http://arxiv.org/abs/hep-th/0008001}{{\tt arXiv:hep-th/0008001}}.

\bibitem{Gubser:2008px}
S.~S. Gubser, ``{Breaking an Abelian gauge symmetry near a black hole
  horizon},'' \href{http://dx.doi.org/10.1103/PhysRevD.78.065034}{{\em Phys.
  Rev.} {\bf D78} (2008)  065034},
\href{http://arxiv.org/abs/0801.2977}{{\tt arXiv:0801.2977 [hep-th]}}.

\bibitem{Hartnoll:2008vx}
S.~A. Hartnoll, C.~P. Herzog, and G.~T. Horowitz, ``{Building a Holographic
  Superconductor},''
  \href{http://dx.doi.org/10.1103/PhysRevLett.101.031601}{{\em Phys. Rev.
  Lett.} {\bf 101} (2008)  031601},
\href{http://arxiv.org/abs/0803.3295}{{\tt arXiv:0803.3295 [hep-th]}}.

\bibitem{Hartnoll:2008kx}
S.~A. Hartnoll, C.~P. Herzog, and G.~T. Horowitz, ``{Holographic
  Superconductors},''
  \href{http://dx.doi.org/10.1088/1126-6708/2008/12/015}{{\em JHEP} {\bf 12}
  (2008)  015},
\href{http://arxiv.org/abs/0810.1563}{{\tt arXiv:0810.1563 [hep-th]}}.

\bibitem{Son:2008ye}
D.~T. Son, ``{Toward an AdS/cold atoms correspondence: a geometric realization
  of the Schroedinger symmetry},''
  \href{http://dx.doi.org/10.1103/PhysRevD.78.046003}{{\em Phys. Rev.} {\bf
  D78} (2008)  046003},
\href{http://arxiv.org/abs/0804.3972}{{\tt arXiv:0804.3972 [hep-th]}}.

\bibitem{Balasubramanian:2008dm}
K.~Balasubramanian and J.~McGreevy, ``{Gravity duals for non-relativistic
  CFTs},'' \href{http://dx.doi.org/10.1103/PhysRevLett.101.061601}{{\em Phys.
  Rev. Lett.} {\bf 101} (2008)  061601},
\href{http://arxiv.org/abs/0804.4053}{{\tt arXiv:0804.4053 [hep-th]}}.

\bibitem{Maldacena:2008wh}
J.~Maldacena, D.~Martelli, and Y.~Tachikawa, ``{Comments on string theory
  backgrounds with non- relativistic conformal symmetry},'' {\em JHEP} {\bf 10}
  (2008)  072,
\href{http://arxiv.org/abs/0807.1100}{{\tt arXiv:0807.1100 [hep-th]}}.

\bibitem{Herzog:2008wg}
C.~P. Herzog, M.~Rangamani, and S.~F. Ross, ``{Heating up Galilean
  holography},'' {\em JHEP} {\bf 11} (2008)  080,
\href{http://arxiv.org/abs/0807.1099}{{\tt arXiv:0807.1099 [hep-th]}}.

\bibitem{Adams:2008wt}
A.~Adams, K.~Balasubramanian, and J.~McGreevy, ``{Hot Spacetimes for Cold
  Atoms},'' \href{http://dx.doi.org/10.1088/1126-6708/2008/11/059}{{\em JHEP}
  {\bf 11} (2008)  059},
\href{http://arxiv.org/abs/0807.1111}{{\tt arXiv:0807.1111 [hep-th]}}.

\bibitem{Nastase:1999cb}
H.~Nastase, D.~Vaman, and P.~van Nieuwenhuizen, ``{Consistent nonlinear K K
  reduction of 11d supergravity on AdS(7) x S(4) and self-duality in odd
  dimensions},'' \href{http://dx.doi.org/10.1016/S0370-2693(99)01266-6}{{\em
  Phys. Lett.} {\bf B469} (1999)  96--102},
\href{http://arxiv.org/abs/hep-th/9905075}{{\tt arXiv:hep-th/9905075}}.

\bibitem{Nastase:1999kf}
H.~Nastase, D.~Vaman, and P.~van Nieuwenhuizen, ``{Consistency of the AdS(7) x
  S(4) reduction and the origin of self-duality in odd dimensions},''
  \href{http://dx.doi.org/10.1016/S0550-3213(00)00193-0}{{\em Nucl. Phys.} {\bf
  B581} (2000)  179--239},
\href{http://arxiv.org/abs/hep-th/9911238}{{\tt arXiv:hep-th/9911238}}.

\bibitem{Freund1980}
P.~G.~O. Freund and M.~A. Rubin, ``{Dynamics of Dimensional Reduction},''
\href{http://dx.doi.org/10.1016/0370-2693(80)90590-0}{{\em Phys. Lett.} {\bf
  B97} (1980)  233--235}.

\bibitem{Gauntlett:2007ma}
J.~P. Gauntlett and O.~Varela, ``{Consistent Kaluza-Klein Reductions for
  General Supersymmetric AdS Solutions},''
  \href{http://dx.doi.org/10.1103/PhysRevD.76.126007}{{\em Phys. Rev.} {\bf
  D76} (2007)  126007},
\href{http://arxiv.org/abs/0707.2315}{{\tt arXiv:0707.2315 [hep-th]}}.

\bibitem{Bremer:1998zp}
M.~S. Bremer, M.~J. Duff, H.~Lu, C.~N. Pope, and K.~S. Stelle, ``{Instanton
  cosmology and domain walls from M-theory and string theory},''
  \href{http://dx.doi.org/10.1016/S0550-3213(98)00764-0}{{\em Nucl. Phys.} {\bf
  B543} (1999)  321--364},
\href{http://arxiv.org/abs/hep-th/9807051}{{\tt arXiv:hep-th/9807051}}.

\bibitem{Liu:2000gk}
J.~T. Liu and H.~Sati, ``{Breathing mode compactifications and supersymmetry of
  the brane-world},''
  \href{http://dx.doi.org/10.1016/S0550-3213(01)00179-1}{{\em Nucl. Phys.} {\bf
  B605} (2001)  116--140},
\href{http://arxiv.org/abs/hep-th/0009184}{{\tt arXiv:hep-th/0009184}}.

\bibitem{Buchel:2006gb}
A.~Buchel and J.~T. Liu, ``{Gauged supergravity from type IIB string theory on
  Y(p,q) manifolds},''
  \href{http://dx.doi.org/10.1016/j.nuclphysb.2007.03.001}{{\em Nucl. Phys.}
  {\bf B771} (2007)  93--112},
\href{http://arxiv.org/abs/hep-th/0608002}{{\tt arXiv:hep-th/0608002}}.

\bibitem{Gauntlett:2009zw}
J.~P. Gauntlett, S.~Kim, O.~Varela, and D.~Waldram, ``{Consistent
  supersymmetric Kaluza--Klein truncations with massive modes},''
  \href{http://dx.doi.org/10.1088/1126-6708/2009/04/102}{{\em JHEP} {\bf 04}
  (2009)  102},
\href{http://arxiv.org/abs/0901.0676}{{\tt arXiv:0901.0676 [hep-th]}}.

\bibitem{Gauntlett:2009dn}
J.~P. Gauntlett, J.~Sonner, and T.~Wiseman, ``{Holographic superconductivity in
  M-Theory},'' \href{http://dx.doi.org/10.1103/PhysRevLett.103.151601}{{\em
  Phys. Rev. Lett.} {\bf 103} (2009)  151601},
\href{http://arxiv.org/abs/0907.3796}{{\tt arXiv:0907.3796 [hep-th]}}.

\bibitem{Gauntlett:2009bh}
J.~P. Gauntlett, J.~Sonner, and T.~Wiseman, ``{Quantum Criticality and
  Holographic Superconductors in M- theory},''
  \href{http://dx.doi.org/10.1007/JHEP02(2010)060}{{\em JHEP} {\bf 02} (2010)
  060},
\href{http://arxiv.org/abs/0912.0512}{{\tt arXiv:0912.0512 [hep-th]}}.

\bibitem{Gubser:2009qm}
S.~S. Gubser, C.~P. Herzog, S.~S. Pufu, and T.~Tesileanu, ``{Superconductors
  from Superstrings},''
\href{http://arxiv.org/abs/0907.3510}{{\tt arXiv:0907.3510 [hep-th]}}.

\bibitem{Cassani:2010uw}
D.~Cassani, G.~Dall'Agata, and A.~F. Faedo, ``{Type IIB supergravity on
  squashed Sasaki-Einstein manifolds},''
\href{http://arxiv.org/abs/1003.4283}{{\tt arXiv:1003.4283 [hep-th]}}.

\bibitem{Gauntlett:2010vu}
J.~P. Gauntlett and O.~Varela, ``{Universal Kaluza-Klein reductions of type IIB
  to N=4 supergravity in five dimensions},''
\href{http://arxiv.org/abs/1003.5642}{{\tt arXiv:1003.5642 [hep-th]}}.

\bibitem{Liu:2010sa}
J.~T. Liu, P.~Szepietowski, and Z.~Zhao, ``{Consistent massive truncations of
  IIB supergravity on Sasaki-Einstein manifolds},''
\href{http://arxiv.org/abs/1003.5374}{{\tt arXiv:1003.5374 [hep-th]}}.

\bibitem{Skenderis:2010vz}
K.~Skenderis, M.~Taylor, and D.~Tsimpis, ``{A consistent truncation of IIB
  supergravity on manifolds admitting a Sasaki-Einstein structure},''
\href{http://arxiv.org/abs/1003.5657}{{\tt arXiv:1003.5657 [hep-th]}}.

\bibitem{Bobev:2010ib}
N.~Bobev, N.~Halmagyi, K.~Pilch, and N.~P. Warner, ``{Supergravity
  Instabilities of Non-Supersymmetric Quantum Critical Points},''
  \href{http://dx.doi.org/10.1088/0264-9381/27/23/235013}{{\em Class. Quant.
  Grav.} {\bf 27} (2010)  235013},
\href{http://arxiv.org/abs/1006.2546}{{\tt arXiv:1006.2546 [hep-th]}}.

\bibitem{Pope:1987ad}
C.~N. Pope and K.~S. Stelle, ``{ZILCH CURRENTS, SUPERSYMMETRY AND KALUZA-KLEIN
  CONSISTENCY},''
\href{http://dx.doi.org/10.1016/0370-2693(87)91487-0}{{\em Phys. Lett.} {\bf
  B198} (1987)  151}.

\bibitem{Cvetic:2000dm}
M.~Cvetic, H.~Lu, and C.~N. Pope, ``{Consistent Kaluza-Klein sphere
  reductions},'' \href{http://dx.doi.org/10.1103/PhysRevD.62.064028}{{\em Phys.
  Rev.} {\bf D62} (2000)  064028},
\href{http://arxiv.org/abs/hep-th/0003286}{{\tt arXiv:hep-th/0003286}}.

\bibitem{Faulkner:2009am}
T.~Faulkner, G.~T. Horowitz, J.~McGreevy, M.~M. Roberts, and D.~Vegh,
  ``{Photoemission 'experiments' on holographic superconductors},''
  \href{http://dx.doi.org/10.1007/JHEP03(2010)121}{{\em JHEP} {\bf 03} (2010)
  121},
\href{http://arxiv.org/abs/0911.3402}{{\tt arXiv:0911.3402 [hep-th]}}.

\bibitem{Gubser:2009dt}
S.~S. Gubser, F.~D. Rocha, and P.~Talavera, ``{Normalizable fermion modes in a
  holographic superconductor},''
\href{http://arxiv.org/abs/0911.3632}{{\tt arXiv:0911.3632 [hep-th]}}.

\bibitem{Ammon:2010pg}
M.~Ammon, J.~Erdmenger, M.~Kaminski, and A.~O'Bannon, ``{Fermionic Operator
  Mixing in Holographic p-wave Superfluids},''
  \href{http://dx.doi.org/10.1007/JHEP05(2010)053}{{\em JHEP} {\bf 05} (2010)
  053},
\href{http://arxiv.org/abs/1003.1134}{{\tt arXiv:1003.1134 [hep-th]}}.

\bibitem{Andrianopoli:1996cm}
L.~Andrianopoli {\em et al.}, ``{N = 2 supergravity and N = 2 super Yang-Mills
  theory on general scalar manifolds: Symplectic covariance, gaugings and the
  momentum map},'' \href{http://dx.doi.org/10.1016/S0393-0440(97)00002-8}{{\em
  J. Geom. Phys.} {\bf 23} (1997)  111--189},
\href{http://arxiv.org/abs/hep-th/9605032}{{\tt arXiv:hep-th/9605032}}.

\bibitem{Duff:1984sv}
M.~J. Duff, B.~E.~W. Nilsson, and C.~N. Pope, ``{THE CRITERION FOR VACUUM
  STABILITY IN KALUZA-KLEIN SUPERGRAVITY},''
\href{http://dx.doi.org/10.1016/0370-2693(84)91234-6}{{\em Phys. Lett.} {\bf
  B139} (1984)  154}.

\bibitem{Martelli:2006yb}
D.~Martelli, J.~Sparks, and S.-T. Yau, ``{Sasaki-Einstein manifolds and volume
  minimisation},'' \href{http://dx.doi.org/10.1007/s00220-008-0479-4}{{\em
  Commun. Math. Phys.} {\bf 280} (2008)  611--673},
\href{http://arxiv.org/abs/hep-th/0603021}{{\tt arXiv:hep-th/0603021}}.

\bibitem{Hitchin19741}
N.~Hitchin, ``Harmonic spinors,''
  \href{http://dx.doi.org/DOI:10.1016/0001-8708(74)90021-8}{{\em Advances in
  Mathematics} {\bf 14} (1974) no.~1, 1 -- 55}.

\bibitem{Pope:1984jj}
C.~N. Pope and N.~P. Warner, ``{TWO NEW CLASSES OF COMPACTIFICATIONS OF d = 11
  SUPERGRAVITY},''
\href{http://dx.doi.org/10.1088/0264-9381/2/1/001}{{\em Class. Quant. Grav.}
  {\bf 2} (1985)  L1}.

\bibitem{Gibbons:2002th}
G.~W. Gibbons, S.~A. Hartnoll, and C.~N. Pope, ``{Bohm and Einstein-Sasaki
  metrics, black holes and cosmological event horizons},''
  \href{http://dx.doi.org/10.1103/PhysRevD.67.084024}{{\em Phys. Rev.} {\bf
  D67} (2003)  084024},
\href{http://arxiv.org/abs/hep-th/0208031}{{\tt arXiv:hep-th/0208031}}.

\bibitem{Liu:2009dm}
H.~Liu, J.~McGreevy, and D.~Vegh, ``{Non-Fermi liquids from holography},''
\href{http://arxiv.org/abs/0903.2477}{{\tt arXiv:0903.2477 [hep-th]}}.

\bibitem{Faulkner:2009wj}
T.~Faulkner, H.~Liu, J.~McGreevy, and D.~Vegh, ``{Emergent quantum criticality,
  Fermi surfaces, and AdS2},''
\href{http://arxiv.org/abs/0907.2694}{{\tt arXiv:0907.2694 [hep-th]}}.

\bibitem{Chen:2009pt}
J.-W. Chen, Y.-J. Kao, and W.-Y. Wen, ``{Peak-Dip-Hump from Holographic
  Superconductivity},''
  \href{http://dx.doi.org/10.1103/PhysRevD.82.026007}{{\em Phys. Rev.} {\bf
  D82} (2010)  026007},
\href{http://arxiv.org/abs/0911.2821}{{\tt arXiv:0911.2821 [hep-th]}}.

\bibitem{Gubser:2010dm}
S.~S. Gubser, F.~D. Rocha, and A.~Yarom, ``{Fermion correlators in non-abelian
  holographic superconductors},''
\href{http://arxiv.org/abs/1002.4416}{{\tt arXiv:1002.4416 [hep-th]}}.

\bibitem{Freedman:1976aw}
D.~Z. Freedman and A.~K. Das, ``{Gauge Internal Symmetry in Extended
  Supergravity},''
\href{http://dx.doi.org/10.1016/0550-3213(77)90041-4}{{\em Nucl. Phys.} {\bf
  B120} (1977)  221}.

\bibitem{Fradkin:1976xz}
E.~S. Fradkin and M.~A. Vasiliev, ``{Model of Supergravity with Minimal
  Electromagnetic Interaction},''. LEBEDEV-76-197.

\bibitem{Romans:1991nq}
L.~J. Romans, ``{Supersymmetric, cold and lukewarm black holes in cosmological
  Einstein-Maxwell theory},''
  \href{http://dx.doi.org/10.1016/0550-3213(92)90684-4}{{\em Nucl. Phys.} {\bf
  B383} (1992)  395--415},
\href{http://arxiv.org/abs/hep-th/9203018}{{\tt arXiv:hep-th/9203018}}.

\end{thebibliography}\endgroup

\end{document}